\documentclass{ametsocV6.1}
\usepackage{hyperref}

\title{Emergence of an Advective Boundary Layer in Monsoon Cross-Equatorial Flow: Scaling, Dynamics, and Idealized Models}

\authors{Rajat Masiwal,\aff{a,b}\correspondingauthor{Rajat Masiwal, rajatm@uchicago.edu} 
Ashwin K Seshadri,\aff{b,c}
	Vishal Dixit,\aff{d}  
}

\affiliation{
    \aff{a}{Department of the Geophysical Sciences, The University of Chicago, Chicago, IL, USA}\\
    \aff{b}{Centre for Atmospheric and Oceanic Sciences, Indian Institute of Science, Bengaluru, India}\\
	\aff{c}{Divecha Centre for Climate Change, Indian Institute of Science, Bengaluru, India}\\
	\aff{d}{Centre for Climate Studies, Indian Institute of Technology Bombay, Mumbai, India}\\
}
\abstract{The conventional Ekman model of the tropical boundary layer neglects nonlinear momentum advection and breaks down near the equator, where Coriolis effects are weak. During South Asian monsoon onset, we identify a dynamical regime transition to an advective boundary layer (ABL). Reanalysis links this transition to a shift in the zonal momentum balance from frictional to meridional-advection control as cross-equatorial flow intensifies, accompanied by increasing local Rossby number and vanishing absolute vorticity, signaling the breakdown of Ekman balance. A scaling analysis shows that this transition occurs when the meridional length scales of geopotential and zonal wind contract such that their product approaches $\phi/f^2$. In the resulting ABL regime, kinetic energy is governed by a balance between its generation and advection, yielding a linear diagnostic relation between meridional geopotential gradient and meridional wind. A simple theoretical model predicts that the sensitivity of this relation is controlled by an advective timescale that equals the inertial timescale ($1/f$) at the transition latitude, where zonal and meridional wind speeds become comparable. Testing this framework in idealized aquaplanet experiments confirms that stronger cross-equatorial pressure gradients and slower planetary rotation rates amplify advective effects and shift the transition latitude poleward. Across experiments, the sensitivity of meridional winds to the geopotential gradient remains tightly linked to $1/f$ at the transition latitude. Together, these results establish the ABL as a distinct dynamical regime, with important implications for monsoon onset, intraseasonal variability, and the representation of tropical boundary layer processes in climate models.} 

\begin{document}

\maketitle
\nolinenumbers

%
\section{Introduction} \label{1}

The tropical atmospheric boundary layer (BL), extending from the Earth's surface to the cloud base (typically 1–2 km), plays a central role in the climate system. It is the primary interface between the atmosphere and the surface, contains the bulk of atmospheric moisture, and mediates surface fluxes and the vertical transport of momentum and energy \citep{garratt1994atmospheric}. Thermodynamic and dynamic variables are generally well-mixed within the BL \citep{mcgauley2004large,johnson2001multiscale}, shaping near-surface winds, convergence, and pressure gradients that organize tropical convection and rainfall\citep{holton1971boundary,lindzen1987role,waliser1994preferred,tomas1999influence,sobel2006boundary,dixit2017role,byrne2019dynamics}. Despite this central role, the dynamical balances governing the tropical boundary layer, particularly near the equator, remain incompletely understood.

The tropical boundary layer is often modeled using a bulk formulation that assumes a steady balance between the Coriolis force, the pressure-gradient force, and surface friction \citep{lindzen1987role,stevens2002entrainment,mcgauley2004large,byrne2019dynamics}, commonly referred to as Ekman balance \citep{ekman1905influence}. Extended formulations also include entertainment at the boundary layer top from the free troposphere \citep{stevens2002entrainment,back2009relationship}. Nonlinear horizontal advection of momentum is typically neglected in these formulations under the assumption that this is comparatively negligible. However, observational and modeling studies increasingly challenge this assumption, demonstrating that under specific large-scale conditions, advective terms can be comparable to, or even dominate, the other terms in the boundary-layer momentum budget \citep{yang2013zonal,dixit2017role,gonzalez2019violation,gonzalez2022rapid,praturi2025meridional}. Moreover, near the equator, the smallness of the Coriolis parameter renders the Ekman balance incomplete, motivating the need for frameworks that explicitly account for additional effects including that of momentum advection.

Previous studies have found that horizontal momentum advection becomes increasingly important in the presence of strong cross-equatorial pressure gradients, leading to a boundary layer quite distinct from the Ekman boundary layer \citep{mahrt1972numerical,tomas1999influence,gonzalez2019violation}. This situation is especially relevant for the South Asian monsoon, where strong cross-equatorial pressure gradients are observed, which drive an intense low-level flow that turns eastward into the Somali jet \citep{tomas1999influence,krishnamurti1976numerical}. Prior observational and boundary-layer modeling studies have argued that nonlinear momentum advection plays a leading-order role in the dynamics of this jet \citep{krishnamurti1979planetary,yang2013zonal,seshadri2022kinetic}. Recent work has further demonstrated that the abrupt acceleration of the Somali jet is associated with a regime transition in the boundary-layer momentum balance, from frictionally controlled Ekman balance to an advectively dominated balance that emerges during the onset of cross equatorial flow \citep{masiwal2023explaining}. We refer to this regime as the advective boundary layer (ABL), which has been identified as a key feature of the low-level flow in the western Indian Ocean especially during the abrupt transition to the boreal summer monsoon. Although the ABL has been identified as a key feature of the seasonal transition, a mechanistic framework for its emergence is currently missing.


To derive mechanistic explanations and elucidate the essential dynamics of atmospheric phenomena, idealized modeling frameworks such as aquaplanet simulations have often been deployed. Aquaplanet configurations retain core features of atmospheric dynamics while simplifying—and in some cases omitting—components of the broader climate system in order to reduce model complexity~\citep{neale2000standard}. These simulations have been used to probe a wide range of tropical phenomena, including the mean meridional circulation and the associated intertropical convergence zone (ITCZ)~\citep{mobis2012factors,dixit2017role,geen2019processes}, tropical cyclones~\citep{merlis2013sensitivity,merlis2019aquaplanet}, and intraseasonal oscillations~\citep{bellon2008instability,maloney2010intraseasonal,das2016low}. Using idealized dry as well as moist simulations, previous studies have also highlighted the importance of nonlinear advective terms in the horizontal momentum budget of the atmospheric boundary layer, across models of varying complexity~\citep{tomas1999influence,schneider2008eddy,toma2010oscillations,dixit2017role,gonzalez2016dynamics,faulk2017effects,byrne2019dynamics,gonzalez2019violation}. Much of this literature examines ITCZ dynamics and its off-equatorial migration by varying the latitude and structure of imposed forcings to drive changes in the Hadley circulation. Within this framework, several studies have noted that boundary-layer dynamics can modulate key ITCZ characteristics—including its latitude, width, and intraseasonal variability~\citep{toma2010oscillations,byrne2019dynamics,faulk2017effects}. Despite these advances, the dynamical conditions under which an advective boundary-layer (ABL) regime emerges within cross-equatorial flow, and how that emergence depends on the spatial structure of forcing, remain incompletely understood.

That idealized simulations can reproduce salient ABL features when the forcing latitude is shifted suggests that the meridional length scale of the forcing, and not only its amplitude, may help set the conditions for an ABL regime to develop dynamically from cross-equatorial flow. This hypothesis is consistent with prior emphasis on the role of cross-equatorial pressure gradients, but a scale analysis that explicitly accounts for the relevant meridional length scales of forcing and response is still lacking. More broadly, it remains unclear whether the ABL behavior seen in idealized models and the observed tropical boundary layer can be reconciled within a unified theory based on scaling arguments. The present study combines observations, scaling theory, and idealized aquaplanet experiments to address this gap.

We show that the boundary-layer transition from Ekman to advective dynamics is governed by changes in the meridional length scales of geopotential and zonal wind. The local Rossby number, defined here as the ratio of advective to Coriolis acceleration, ($Ro \equiv -\frac{\zeta}{f}$), provides a natural diagnostic of this transition: it encapsulates the growing importance of nonlinear advection as horizontal shear intensifies and as the relevant horizontal length scales shorten amidst the anisotropy of horizontal length scales between zonal and momentum directions. This role of $Ro$ in atmospheric boundary-layer (ABL) dynamics is distinct from its use in previous studies as a nondimensional measure of eddy influence on the Hadley circulation and its connection to angular-momentum conservation~\citep{walker2005response,schneider2006general,schneider2008eddy,bordoni2008monsoons}. In the ABL regime, as $Ro$ becomes large, the zonal momentum budget simplifies at leading order: meridional advection of zonal momentum balances the pressure-gradient force, while friction becomes comparatively small. This balance yields a linear diagnostic relation between meridional wind and the meridional geopotential gradient, with a proportionality set by an advective timescale that scales as $f^{-1}$.

To test the robustness and generality of this framework, we conduct a suite of idealized simulations using a moist aquaplanet general circulation model with zonally symmetric sea surface temperature (SST) forcing. By systematically varying the latitude of the SST maximum, we control the meridional length scale of geopotential, and thereby the strength and spatial structure of the cross-equatorial pressure gradient,  modulating the onset and structure of the advective ABL regime. We additionally vary the planetary rotation rate to assess how the sensitivity of meridional winds, and therefore nonlinear momentum advection, depends on the inertial timescale.

This paper is organized as follows. Section \ref{5.2} describes the data, diagnostics, and model configuration. Section \ref{5.3} presents observational evidence for the advective boundary layer in the Somali jet region and develops the associated scaling theory, followed by validation in an idealized GCM. Section \ref{5.4} summarizes the main findings and discusses broader implications.

\section{Data and Method} \label{5.2}
\subsection{Data and domain} \label{5.2.1}
We use the European Centre for Medium-Range Weather Forecasts (ECMWF) ERA5 reanalysis \citep{hersbach2020era5}. Data for horizontal and vertical velocity, vorticity, and geopotential are downloaded at a horizontal resolution of $0.25^\circ\times0.25^\circ$. Daily mean values of each variable are computed by averaging 6-hourly data (at 00,06,12 and 18 UTC). Since our focus is on the boundary layer, all the analysis is performed at 900 hPa for the period of 1979-2020. Daily rainfall values over the core monsoon zone (CMZ) are estimated using rain-gauge based gridded daily rainfall data from India Meteorological Department~\citep{rajeevan2006high}.

During the boreal summer, the most intense cross-equatorial flow in the tropical boundary layer is observed in the western Arabian Sea. This flow gives rise to the Somali jet, which is responsible for a significant fraction of moisture advected toward the Indian subcontinent. For the Somali jet, \citet{masiwal2023explaining} reported that the advective terms become the most important in the ``N.Eq.'' region (50$^\circ$E-60$^\circ$E, 2$^\circ$N-10$^\circ$N) and moreover are much larger in magnitude than the frictional dissipation there. Therefore, the observational analysis of this study will mainly focus on this region (see Fig. S1). 

\subsection{Onset of cross-equatorial flow} \label{5.2.2}
For the composite analyses of the various terms examined in this study, we first select a reference day for each year defined by the condition that following this day the 900 hPa flow becomes persistently cross-equatorial from the southern to northern hemisphere. This reference day is identified as the day on which the meridional wind ($v$) averaged over the ``N.Eq.'' region first becomes southerly ($v>0$) and remains so during each of the next 15 days. The reference day defines the onset of the cross-equatorial flow during the seasonal transition within the boundary layer.

\subsection{Local Rossby number ({$R_o$}) and absolute vorticity ({$\eta$})} \label{5.2.3}
Various studies have used absolute vorticity ({$\eta$}) and the local Rossby number ({$R_o$}) to discuss either the inertial instability or the cross-equatorial flow in the boundary layer \citep{tomas1997role,tomas1999influence,toma2010oscillations,schneider2008eddy,masiwal2023explaining}. Given $R_o$ defined as the ratio of relative vorticity ($\zeta$) to planetary vorticity ($f$), i.e., $R_{o} = -\zeta/f$, the absolute vorticity $\eta$ defined as the sum of relative vorticity and planetary vorticity, i.e., $\eta = \zeta+f$ can be written as
\begin{equation}
    \eta = f(1-R_{o})
    \label{eq5.1}
\end{equation}
From equation \ref{eq5.1}, it is clear that where $R_{o} =1$, $\eta=0$. Since these two conditions are equivalent, we will use them interchangeably throughout this study, as we examine the evolution of the contour of $\eta=0$ into the northern hemisphere. 

\subsection{Model Configuration} \label{5.2.4}
Idealized modeling experiments are performed to understand the factors controlling the advective boundary layer, using the National Center for Atmospheric Research (NCAR) community atmosphere model (CAM) version 4 \citep{gent2011community}. CAM is the atmospheric component of the Community Earth System Model (CESM 2.1.3; \citep{danabasoglu2020community}). The model is run in the aquaplanet setting with a finite volume dynamical core at a horizontal resolution of 1.9$^\circ$ latitude $\times$ 2.5$^\circ$ longitude and 26 levels in the vertical. The model is forced with a fixed zonally symmetric sea surface temperature (SST) distribution. The meridional SST profile is specified following Neale and Hoskins (2000)~\citep{neale2000standard}, with SST (in $^\circ$C) defined as a function of latitude ($y$) by:

\begin{equation}
SST(y)=
    \begin{cases}
        0 & \text{if } y > \frac{\pi}{3}\\
        27\left(1-\sin^2\left(\frac{\pi}{2} \left[\frac{y-y_{0}}{\frac{\pi}{3} - y_{0}}\right]\right)\right) & \text{if } \frac{\pi}{3} \geq y \geq y_{0}\\
        27\left(1-\sin^2\left(\frac{\pi}{2} \left[\frac{y-y_{0}}{\frac{\pi}{3} + y_{0}}\right]\right)\right) & \text{if } y_{0} > y \geq -\frac{\pi}{3}\\
        0 & \text{if } y < -\frac{\pi}{3}
    \end{cases}
    \label{eq5.2}
\end{equation}

To mimic boreal summer-like conditions with warmer SST in the northern hemisphere and a cross-equatorial pressure gradient, we vary the latitude of SST maximum ($y_{0}$ in equation \ref{eq5.2}) in our experiments. Specifically, $y_{0}$ is varied from 0 to 25$^\circ$N across experiments. Since the aquaplanet GCM can require up to 30 to 60 days for spin-up \citep{neale2000standard}, the model is run for one year in each experiment, and the last 160 days of the run are analyzed for the study. All the steady-state fields are obtained after averaging over the last 160 days.

\subsection{Horizontal momentum budget:}
For a hydrostatically balanced atmosphere, the horizontal momentum equations can be represented in pressure coordinates as follows:
\begin{equation}\label{eq3}
    \frac{\partial{u}}{\partial{t}} + 
    u\frac{\partial{u}}{\partial x} +
    v\frac{\partial{u}}{\partial{y}} +
    \omega\frac{\partial{u}}{\partial{p}} -
    fv =
    -\frac{\partial{\phi}}{\partial{x}}+
    F_X
\end{equation}
\begin{equation}\label{eq4}
    \frac{\partial{v}}{\partial{t}} + 
    u\frac{\partial{v}}{\partial x} +
    v\frac{\partial{v}}{\partial{y}} +
    \omega\frac{\partial{v}}{\partial{p}} +
    fu =
    -\frac{\partial{\phi}}{\partial{y}}+
    F_Y
\end{equation}

Here, $u, v, \omega$ are the wind components, $f$ is the Coriolis parameter (or planetary vorticity), and $\phi$ is the geopotential. $F_X$ and $F_Y$ are the accelerations due to all subgrid scale processes in the zonal and meridional directions, respectively. Each term, except for $F_X$ and $F_Y$, in equations \ref{eq3} and \ref{eq4} is computed from daily data in both observations and model simulations. $F_X$ and $F_Y$ are inferred as the residue of the budget and will be labeled as \textit{Res.} in the next section. Moreover, since the focus is primarily on the boundary layer, we expect this term to be dominated by frictional or dissipative processes. 

\section{Results} \label{5.3}
\subsection{Observed momentum balances in the advective boundary layer}\label{5.3.1}
\begin{figure}[!htb]
\centering
\noindent\includegraphics[width=\textwidth]{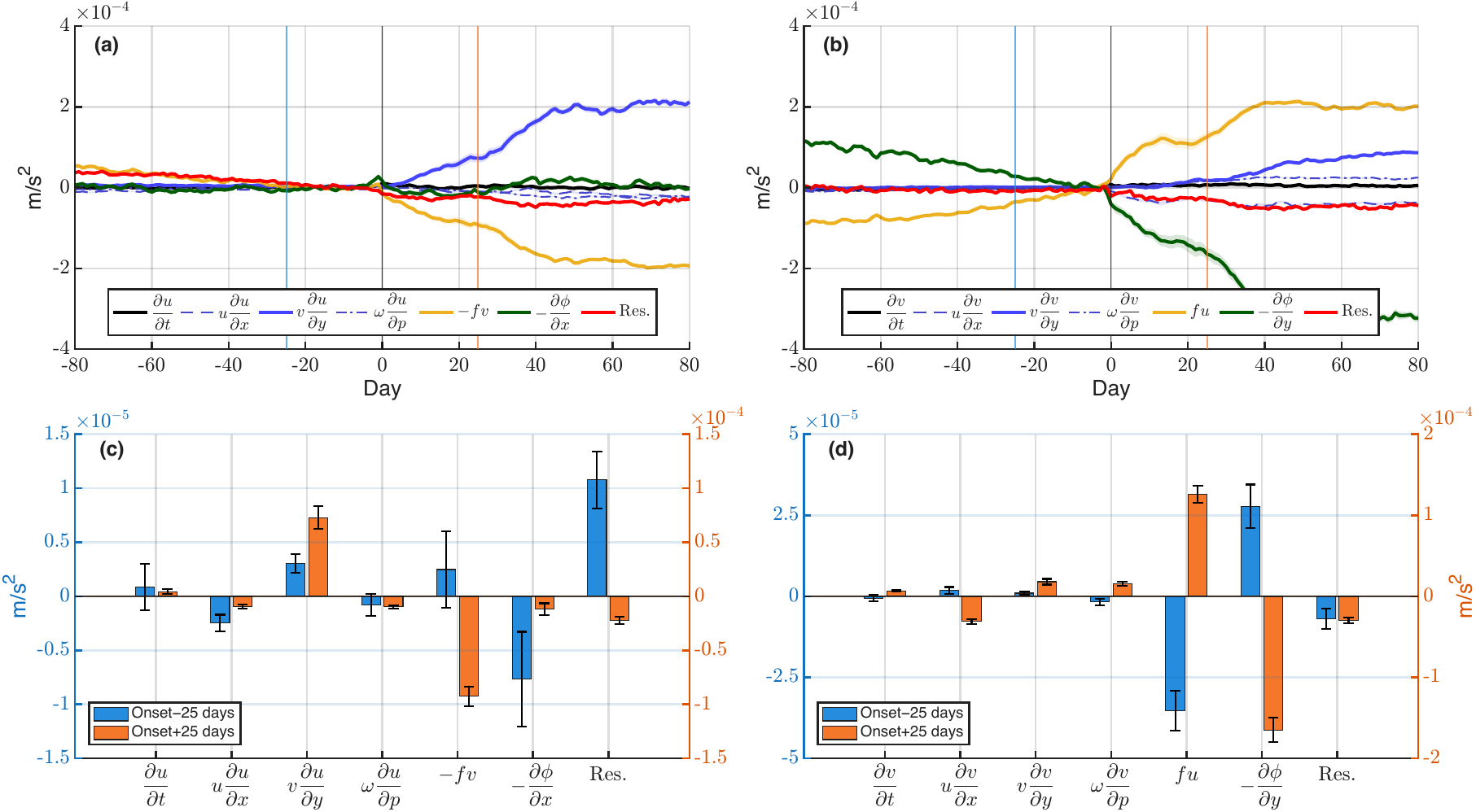}
\caption{Composite evolution of 900 hPa (a) zonal momentum and (b) meridional momentum terms during the seasonal transition. Day 0 corresponds to the onset of cross-equatorial flow (see section \ref{5.2.2} for the definition). Contribution of each term to the (c) zonal and (d) meridional momentum budget at 900 hPa 25 days before (blue) and after (red) the onset of cross-equatorial flow. Note that the scale of terms before (on the left) and after the onset (on the right) is different for both panels. The error-bars indicate the standard error for each term. All the terms are averaged over the ``N.Eq.'' region (50$^\circ$E-60$^\circ$E, 2$^\circ$N-10$^\circ$N).}
\label{fig1}
\end{figure}

We begin by examining the evolution of the horizontal momentum balance in the boundary layer during the onset of cross-equatorial flow. Prior to onset, when the low-level flow in the ``N.Eq.'' region (see section \ref{5.2}) is easterly at 900 hPa (Fig. S1a), the dominant contribution to the zonal momentum balance is the residual, which can be interpreted as friction providing a eastward acceleration that opposes the easterly flow (Fig. \ref{fig1}a,c). This frictional term is initially balanced by the Coriolis term ($-fv$), and closer to onset (day -25) friction is increasingly balanced by the zonal geopotential gradient (Fig. \ref{fig1}c). Together, these balances are characteristic of an Ekman boundary layer \citep{ekman1905influence,stevens2002entrainment}. 

Following onset, as the cross-equatorial southerly flow intensifies (Fig. S1c), the magnitude of the Coriolis term ($-fv$) increases substantially. Friction alone is no longer sufficient to balance this increase, and the meridional advection of zonal momentum ($v\frac{\partial{u}}{\partial{y}}$) grows to leading order, counteracting the Coriolis term and maintaining the zonal momentum balance (Fig. \ref{fig1}a,c). While the magnitudes of the meridional advection and Coriolis terms increase steadily after the onset, the frictional contribution remains approximately constant (with the change in sign as the flow changes from easterly to westerly). The onset therefore marks a clear dynamical transition from a frictionally controlled Ekman regime to an advectively dominated boundary layer, in which meridional advection of zonal momentum plays a central role. Note that the mechanism of this breakdown of Ekman boundary layer in this region is different from that reported in East Pacific, where meridional advection of meridional momentum is the primary term leading to a non-Ekman boundary layer\citep{gonzalez2019violation,gonzalez2022rapid}.

In contrast, for the western Arabian sea, the meridional momentum balance undergoes a much weaker transition.  Although meridional advection of meridional momentum ($v\frac{\partial{v}}{\partial{y}}$) increases after onset, it never becomes dominant, and the balance stays close to geostrophic throughout the period (Fig. \ref{fig1}b,d). This persistence of geostrophic balance in the meridional direction is a manifestation of the asymmetry between the zonal and meridional momentum equations during the transition to the advective boundary layer.

Clearly, the onset of cross-equatorial flow coincides with a qualitative regime change in the boundary-layer momentum balances. Consistent with \citet{masiwal2023explaining}, this transition is accompanied by a rapid increase in the local Rossby number $R_{o}$. The composite evolution shows that $R_{o}$ begins to rise sharply from a small value at day 0 and exceeds 1 within approximately 20 days (Fig. S2a). An equivalent change but of opposite sign (see equation \ref{eq5.1}) is also evident in the evolution of absolute vorticity ($\eta=\zeta+f$), which decreases from a positive value at day 0 to negative values in about 25 days (Fig. S2b).

To quantify this regime shift, we compare the magnitudes of individual terms in the zonal and meridional momentum budgets 25 days before (day -25) and 25 days after (day 25) the onset of sustained cross-equatorial flow (Fig. \ref{fig1}c,d). Prior to onset (day -25), when $R_o\approx -0.1$, the zonal geopotential gradient is primarily balanced by friction, with small contribution from advective terms. By day 25 after onset, as $R_o$ approaches unity, the zonal momentum budget is instead closed by a near balance between the Coriolis term and meridional advection, with friction contributing only weakly (Fig. \ref{fig1}c). In contrast, the meridional momentum balance remains geostrophic at both time slices, with the meridional geopotential gradient balanced by the Coriolis force (Fig. \ref{fig1}d).

Together, these results demonstrate that the transition to the advective boundary layer is tightly linked to $R_{o}$ approaching unity ($R_{o}\rightarrow1$ or equivalently $\eta\rightarrow0$), and manifests as a selective breakdown of the Ekman balance that previously occurred in the zonal momentum equation, while geostrophic balance in the meridional momentum equation is largely preserved. In the next section, we build on this observational evidence by simplifying the momentum equations, performing a scale analysis, and identifying the theoretical constraints on horizontal flow as $\eta$ approaches zero. 

\subsection{Theoretical Considerations}\label{3.2}

\subsubsection{Evolution of length-scales during the seasonal transition} \label{3.2.1}
Two salient features of the seasonal transition emerging from the observational analysis are: 

\begin{enumerate}
  \item the zonal wind remains largely in geostrophic balance throughout the transition (Fig. \ref{fig1}b,d), i.e., 
\begin{equation}
     fu = -\frac{\partial{\phi}}{\partial{y}}
  \label{eq5.3}
\end{equation}
  
  \item the absolute vorticity ($\eta = \zeta+f$) over the ``N.Eq.'' region evolves from positive values prior to onset of cross-equatorial flow to near-zero values around the transition and becomes negative thereafter (Fig. S2b). 
\end{enumerate}
 
We now combine these observational constraints to demonstrate that the seasonal transition into the advective boundary layer can be characterized in terms of changes in the meridional length scales of the geopotential field and the zonal wind. 

Let $B_{u}$ and $B_{\phi}$ denote the characteristic meridional length scales of variation of zonal wind and geopotential, respectively. Characteristic amplitudes of variables are denoted by curly braces $\{\cdot\}$. From the geostrophic balance in  \ref{eq5.3}, the characteristic zonal wind amplitude satisfies 
\begin{equation}
     \{u\} = \frac{\{\phi\}}{fB_{\phi}}
  \label{eq5.4}
\end{equation}
Horizontal variations are strongly anisotropic in the tropics, with  zonal length scales much larger than meridional length scales, and consequently the relative vorticity is largely dominated by the meridional shear of the zonal wind (Fig. S3). We therefore approximate $\zeta = (\frac{\partial{v}}{\partial{x}}-\frac{\partial{u}}{\partial{y}}) \approx -\frac{\partial{u}}{\partial{y}}$.

We now consider three stages of the seasonal transition.

\textbf{Before onset:} Prior to onset, absolute vorticity is positive, so $\eta = \zeta+f > 0$, which implies $f>\frac{\partial{u}}{\partial{y}}$. Invoking the characteristic length scale $B_u$ for $u$ and substituting equation \ref{eq5.4}, this condition becomes

\begin{equation}
     B_{\phi}B_{u} > \frac{\{\phi\}}{f^2}
  \label{eq5.5}
\end{equation}

This inequality corresponds to a regime in which meridional gradients are relatively weak and the Ekman balance is dynamically permitted.

\textbf{At onset:} At the transition ($\approx$ day 25), absolute vorticity approaches zero, $\eta = \zeta+f \approx 0$, yielding the threshold condition
\begin{equation}
     B_{\phi}B_{u} \sim \frac{\{\phi\}}{f^2}
  \label{eq5.6}
\end{equation}

This condition marks the threshold at which the Ekman balance breaks down and advective effects become comparable to Coriolis effects.

\textbf{After onset:} Following onset, absolute vorticity becomes negative and the flow acquires a monsoon-like structure (cross-equatorial flow that turns eastward in the northern hemisphere). In this regime,
\begin{equation}
     B_{\phi}B_{u} < \frac{\{\phi\}}{f^2}
  \label{eq5.7}
\end{equation}

indicating that sufficiently sharp meridional gradients favor an advectively dominated boundary layer.

Equations \ref{eq5.5}-\ref{eq5.7} thus provide the conditions of the seasonal transition of the boundary layer. Since variations in geopotential amplitude remain modest through the season, it is primarily the contraction of meridional length scales and thereby their product that controls the boundary layer transition. 

To test this scaling, we examine the daily evolution of $B_{\phi}B_{u}$ and $\phi/f^2$. The length scales $B_{\phi}$ and $B_{u}$ are estimated by fitting an exponential profile (e.g., $\phi = \phi_{o}\exp(\frac{y}{B_{\phi}})$) to the daily meridional structure of $\phi$ and $u$ between 0-10$^\circ$N using least-squares regression. 

\begin{figure}[!htb]
\centering
\noindent\includegraphics[width=\textwidth]{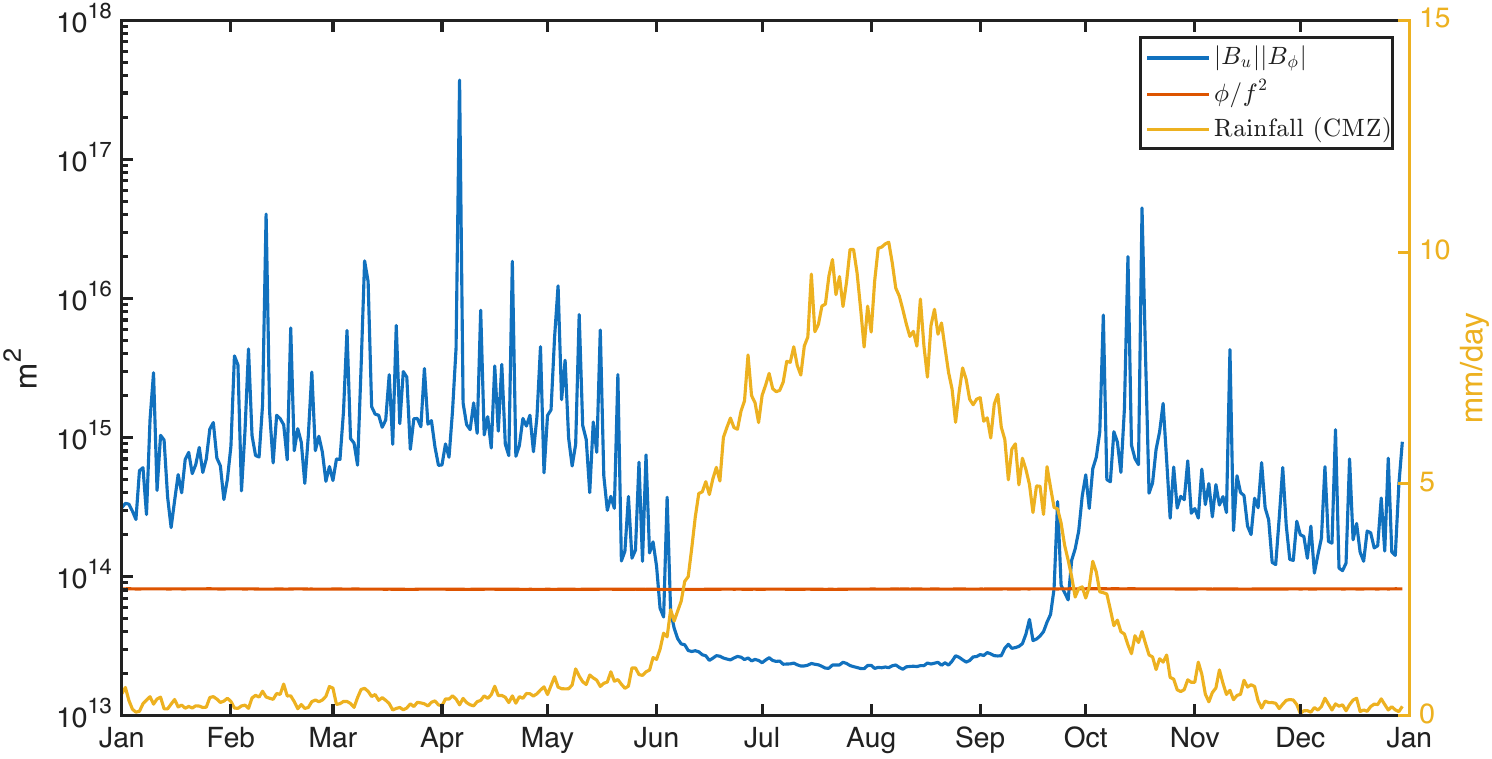}
\caption{Evolution of daily climatology (1979-2020) of $B_{\phi}B_{u}$ and $\phi/f^2$ over the ``N.Eq.'' region. The length scales are estimated using an exponential fit. The vertical dashed line indicates the day when $\eta$ changes sign from positive to negative. The yellow line represents the daily climatology (1979-2020) of rainfall (in $mm/day$) averaged over the core monsoon zone (CMZ) based on India Meteorological Department (IMD) data.}
\label{fig2}
\end{figure}

The daily climatology of $B_{\phi}B_{u}$ and $\phi/f^2$ over the ``N.Eq.'' region shows close agreement with the theoretical scaling. Well before onset, $B_{\phi}B_{u}$ is much larger than $\phi/f^2$. Near the the climatological monsoon onset date (June 1), $B_{\phi}B_{u}$ decreases below $\phi/f^2$ (Fig. \ref{fig2}), coinciding with enhanced rainfall over the core monsoon zone of India. 

Composite evolution relative to the onset date of persistent cross-equatorial flow further corroborates this behavior. As absolute vorticity decreases toward zero around day 0 (Fig. S2b), $B_\phi B_u$ falls below $\phi/f^2$ and remains small as $\eta$ becomes negative (Fig. S4). Afterwards, $B_{\phi}B_{u}$ declines, satisfying the inequality in equation \ref{eq5.7}. The condition where $B_\phi B_u < \phi/f^2$ also qualitatively captures monsoon season length, both in climatology (Fig. \ref{fig2}) and in individual years (Fig. S5), broadly characterizing the period during which South Asian monsoon rainfall is enhanced.

The same contraction of meridional length scales that drives $\eta\rightarrow 0$ also implies a characteristic export time for kinetic energy by meridional advection. In the ABL, kinetic energy generated locally through cross-isobaric flow is removed primarily by meridional advection over the length scale $B_u$ characterizing meridional shear of zonal winds. This motivates an advective timescale of order $\tau \sim B_u/\{v\}$, whose dependence on $f$ emerges naturally at the transition when $\eta\approx 0$. Below, this is made explicit and linked to a diagnostic KE model that was previously introduced by \citet{masiwal2023explaining} to account for abrupt onset of the Somali jet.

\subsubsection{Advective timescale and the relationship between $v$ and $\frac{\partial{\phi}}{\partial{y}}$}\label{3.2.2}

\citet{masiwal2023explaining} introduced a simple one-dimensional framework to account for kinetic energy (KE) evolution within the advective boundary layer during the South Asian monsoon. This framework accounts for the observation that KE undergoes rapid nonlinear intensification, a direct effect of meridional wind $v$ becoming approximately proportional to the local meridional geopotential gradient in the high $R_o$ regime \citep{masiwal2023explaining}. Here we link the empirically estimated constant of proportionality in \citet{masiwal2023explaining} to the length-scale constraints in Section~\ref{3.2.1} as well as the advective timescale $\tau$ that controls KE export.

\textbf{(i) Advective timescale from meridional length scales.}
In the ABL, the dominant KE export is by meridional advection of zonal KE, consistent with the observed dominance of $v\,\partial u/\partial y$ in the zonal momentum balance. Consider the zonal component of kinetic energy,
$KE_u \equiv u^2/2$.
Meridional advection exports $KE_u$ at a rate
\[
v\,\frac{\partial KE_u}{\partial y}
= v\,u\,\frac{\partial u}{\partial y}.
\]
Using the characteristic meridional shear length scale $B_u$,
$\left\{\frac{\partial u}{\partial y}\right\}\sim \{u\}/B_u$,
we obtain the characteristic export rate
\[
\left\{ v\,\frac{\partial KE_u}{\partial y}\right\}
\sim \{v\}\,\{u\}\,\frac{\{u\}}{B_u}
= \frac{\{v\}\,\{u\}^2}{B_u}.
\]
In a quasi-steady ABL, local KE generation is balanced primarily by this export, implying an effective decay timescale $\tau$ defined by
\[
\frac{\{KE_u\}}{\tau}
\sim \left\{ v\,\frac{\partial KE_u}{\partial y}\right\}.
\]
Since $\{KE_u\}\sim \{u\}^2/2$, this yields the advective timescale
\begin{equation}
\tau \sim \frac{B_u}{\{v\}},
\label{eq5.13a}
\end{equation}
up to an order-one factor.
\footnote{A factor of 2 that appears in intermediate scaling estimates disappears once one recognizes that the advective timescale $\tau$ in the kinetic energy model of \citep{masiwal2023explaining} is defined in terms of advection of the total horizontal kinetic energy, $KE=(u^{2}+v^{2})/2$, rather than zonal kinetic energy alone. At the transition latitude, where the flow satisfies $u\approx v$, this definition removes the factor of 2 appearing and yields $\tau \approx 1/f$.}

During the transition, $\eta\approx 0$ implies $f \approx \partial u/\partial y$ (because $\eta=\zeta+f\approx -\partial u/\partial y + f$). Therefore $\{u\}/B_u \sim f$, i.e.,
\begin{equation}
B_u \sim \frac{\{u\}}{f}.
\label{eq5.13b}
\end{equation}
Moreover, by definition of the transition, the flow shifts from predominantly meridional to predominantly zonal so that $\{v\}\sim \{u\}$, which is made explicit below. Substituting \eqref{eq5.13b} into \eqref{eq5.13a} then gives the key scaling
\begin{equation}
\tau \sim \frac{1}{f}
\qquad \text{(evaluated at the transition latitude)}.
\label{eq5.13c}
\end{equation}

Thus, the length-scale condition for $\eta\rightarrow 0$ not only describes the Ekman-ABL transition, but also implies an emergent KE export timescale set by the inertial period.

\textbf{(ii) Connection to the one-dimensional KE model and the $v$-$\partial\phi/\partial y$ relation.}
With this scaling in mind, we now revisit the KE model of \citep{masiwal2023explaining}, written as
\begin{equation}
	\frac{d(KE)}{dt} = KE_{gen}(t) - \frac{KE}{\tau},
 \label{eq5.13}
\end{equation}
where $\tau$ represents the effective export time of horizontal KE by horizontal advection. The generation term arises from pressure work that converts available potential energy into kinetic energy through the horizontal flow across isobars, $KE_{\mathrm{gen}}=-\vec{U}_H\cdot\nabla_H\phi$. 

Within the advective boundary layer, the KE tendency is small compared to KE generation so that a quasi-steady balance applies between KE generation and its decay at an approximately constant advective timescale. Moreover, KE generation in the ``N.Eq.'' region is dominated by the meridional component of the flow and may be approximated as $KE_{gen} \approx -v\,\frac{\partial{\phi}}{\partial{y}}$. Using the geostrophic balance for the zonal wind from equation \eqref{eq5.3}, this expression can equivalently be written as $KE_{gen} \approx vfu$. Imposing the steady-state limit of equation \eqref{eq5.13} and substituting for $KE$ and $KE_{gen}$ yields
\begin{align}
    \frac{KE}{\tau}-KE_{gen} = 0 \nonumber \\
    \frac{u^2+v^2}{2\tau}-vfu = 0 \nonumber \\
    v^2-2v\tau fu+u^2 = 0,
 \label{eq5.14}
\end{align}
which is a quadratic equation for $v$ in terms of $u$ and $\tau$. Its solution is
\[
v = f\tau\,u \pm u\,\sqrt{(f\tau)^2 - 1}.
\]

Near the transition latitude, the flow shifts from being predominantly meridional to predominantly zonal with $u\approx v$ \citep{schneider2008eddy}. Applying this condition to equation \eqref{eq5.14} yields $f\tau \approx 1$, consistent with the order estimate \eqref{eq5.13c} and implying $\tau \approx 1/f$ at the transition latitude.

This result provides a physical interpretation of the timescale $\tau$ inferred by \citet{masiwal2023explaining}, who estimated $\tau\approx10.3$~h from reanalysis, corresponding closely to the inertial timescale at $\sim 10^\circ$N. The observed latitude at which $R_o=1$, averaged over the ``N.Eq.'' longitude band, lies near $8^\circ$N, consistent with this interpretation. The remaining discrepancy likely reflects departures from the assumptions of a strictly constant advective timescale and negligible zonal KE generation, as well as the idealization that friction is negligible. 

Finally, the condition $u\approx v$, together with geostrophic balance for the zonal wind (equation \eqref{eq5.3}), directly yields a linear diagnostic relation between the meridional wind and the geopotential gradient:
\begin{equation}
    v \approx -\frac{1}{f}\frac{\partial{\phi}}{\partial{y}},
    \label{eq5.15}  
\end{equation}
which, at the transition latitude where $\tau \approx 1/f$, may be written equivalently as
\begin{equation}
    v \approx -\tau\frac{\partial{\phi}}{\partial{y}}.
    \label{eq5.16}  
\end{equation}
The proportionality coefficient inferred from observations in the ``N.Eq.'' region after the development of ABL (day 26 to day 80) agrees qualitatively with that obtained from equation \eqref{eq5.15}(Fig. \ref{fig4}a). This supports the interpretation of this linear relationship as a consequence of an emergent advective export timescale that is established during the length-scale contraction associated with $\eta\rightarrow 0$. The advective timescale ($\tau$) as inferred from the linear fit is $-3.5\times10^4$~s (~9.7 hrs) (Fig. \ref{fig4}a). This is unsurprisingly much faster than the Rayleigh friction timescale, which is estimated to be greater than a day over the oceanic boundary layer \citep{deser1993diagnosis,yang2013zonal}. The value of slope ($\tau$) in Fig. \ref{fig4}a corresponds to latitude of $11.3^\circ$N, whereas the latitude where $\eta$ vanishes is around $8^\circ$N. These differences could arise from our neglect of frictional term while deriving equation \eqref{eq5.15}. 

Below, we examine whether the same momentum balances and associated $v$---$\frac{\partial{\phi}}{\partial{y}}$ relation arise in a moist general circulation model under controlled forcing.

\subsection{Advective boundary layer in an aquaplanet}\label{5.3.3}
\subsubsection{Response of the low-level flow to a northward shift of the SST maximum}
We first examine how advective boundary layer dynamics and the flow respond to changing cross-equatorial pressure gradients using a set of idealized moist aquaplanet simulations. In these experiments, cross-equatorial pressure gradients  are imposed by systematically shifting the latitude of the prescribed SST maximum ($y_{0}$ in equation \ref{eq5.2}) northward from the equator. This setup isolates the role of large-scale thermal forcing in the emergence of the advective boundary-layer dynamics, independent of any zonal asymmetries introduced by land-sea contrasts or orography. 

When the SST maximum is located at the equator ($\mathrm{SST_{EQ}}$ experiment), the meridional profile of surface pressure is symmetric about the equator, and no cross-equatorial pressure gradient is present (Fig. S6). As a result, the lower tropospheric flow at 900 hPa winds consists primarily of easterlies (Fig. S7a), and the ITCZ is located at the equator, representative of an equinoctial climate state. Consistent with this hemispheric symmetry, zero absolute vorticity (magenta contour in Fig. S7) also lies close to the equator. 

As the SST maximum is displaced northward, meridional asymmetry develops and cross-equatorial surface pressure gradients are generated, which intensify for increasing $y_{0}$ (Fig. S6c). Given the absence of orography in the aquaplanet GCM, these surface pressure gradients closely approximate geopotential gradients in the lower troposphere \citep{seshadri2022kinetic}. As the maximum SST is displaced northward, a cross-equatorial flow emerges even with modest geopotential gradients at the equator, indicating a sensitive dynamical response of the boundary layer to imposed thermal forcing (Fig. S7c).

This flow is even stronger for experiments with larger gradients at the equator. For experiment with SST maxima at $20^\circ$N and $25^\circ$N, the cross-equatorial flow strengthens considerably and turns eastward under the influence of the Coriolis force, forming a low-level westerly jet near 15$^\circ$N (Fig. S7e,f). This structure resembles the Somali jet observed during boreal summer. The eastward turning of the winds reflects both the strengthening Coriolis deflection with latitude and the persistence of negative meridional geopotential gradients up to 15$^\circ$N (Fig. S6c).

 Accompanying the development of cross-equatorial flow, the ITCZ shifts northward, and the $\eta=0$ contour migrates from the equator into the northern hemisphere (Fig. S7). This displacement is similar to observed seasonal evolution of absolute vorticity in the South Asian monsoon region and signals the development of an advectively dominated boundary layer regime. The appearance of these features in a zonally symmetric aquaplanet suggests that neither land-sea contrast nor regional topography are essential to the advective boundary layer and associated cross-equatorial flow.  
 
 Motivated by these circulation changes that resemble the observed advective boundary layer, we next examine the boundary layer momentum budgets in each experiment to diagnose the dominant balances and assess whether the flow indeed transitions from Ekman to advective dynamics.

\subsubsection{Boundary layer momentum budget}
\begin{figure}[!htb]
\centering
\noindent\includegraphics[width = \textwidth]{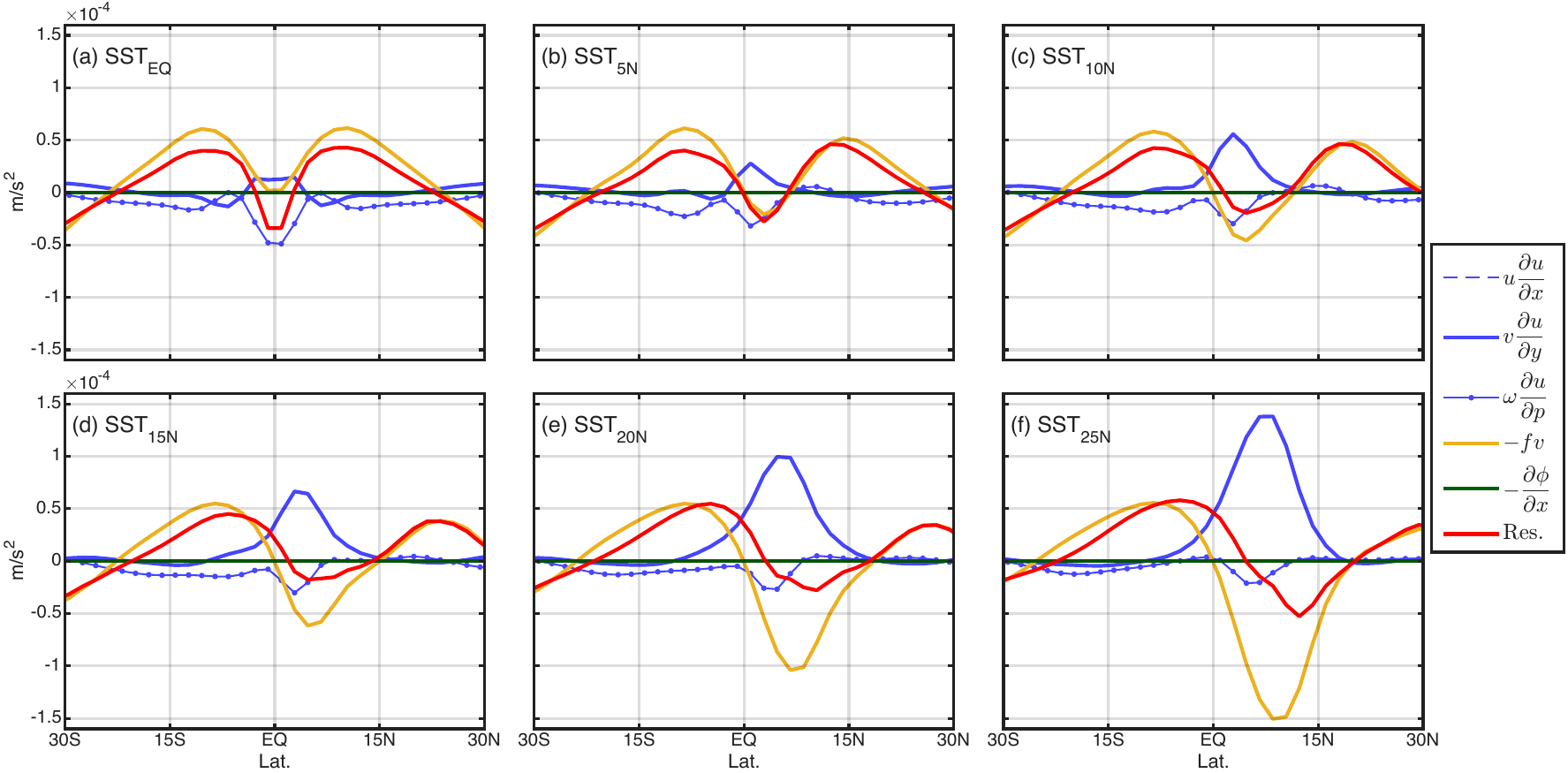}
\caption{Meridional profile of zonal momentum terms for different SST maximum experiments.}
\label{fig3}
\end{figure}

To diagnose the boundary-layer dynamics underlying the circulation changes described above, we examine the zonal and meridional momentum budgets at 900 hPa in the aquaplanet simulations. Fig. \ref{fig3} shows the meridional profiles of the time-averaged zonal momentum term for different SST maximum experiments. For the equatorial SST maximum case ($\mathrm{SST_{EQ}}$), the zonal momentum balance is characteristic of an Ekman boundary layer. Away from the equator, the Coriolis term ($-fv$) is largely balanced by friction, while very close to the equator - where the Coriolis parameter vanishes - friction is balanced by the vertical advection of zonal momentum ($\omega\frac{\partial{u}}{\partial{p}}$) (Fig. \ref{fig3}a). Meridional advection of zonal momentum remains weak throughout the domain, consistent with the absence of a sustained cross-equatorial flow. 

As the SST maximum is displaced northward, introducing a cross-equatorial pressure gradient, the zonal momentum balance undergoes a systematic transition. The contribution of meridional advection of zonal wind ($v\frac{\partial{u}}{\partial{y}}$) increases sensitively north of the equator and becomes comparable to, and eventually balances, the Coriolis term. The contribution of friction to the overall budget becomes comparatively small north of the equator. The transition to such a balance, between Coriolis acceleration and meridional advection of zonal momentum, is characteristic of the development of the advective boundary layer identified in observations of the western Arabian Sea during monsoon onset (Fig. \ref{fig1}a,c). That the same balance emerges in this zonally symmetric and idealized setting demonstrates that strong cross-equatorial pressure gradients alone are sufficient to drive the transition from Ekman to advective dynamics. As the model is zonally symmetric, the contribution of zonal gradients is comparatively small. 

In contrast, the meridional momentum budget remains close to geostrophic balance across all experiments, particularly beyond a few degrees away from the equator (Fig. S8). Only near the equator, and primarily in experiments with a well-developed advective boundary layer, do small deviations to this balance arise. These deviations arise from meridional advection of the meridional wind as well as friction, but the geostrophic balance remains a reasonable first approximation. This persistent asymmetry, with advective dominance in the zonal momentum equation alongside the maintenance of near-geostrophic balance in the meridional direction, is also characteristic of the balance diagnosed in the reanalysis data  (Fig. \ref{fig1}b,d).

These results demonstrate that the emergence of an advective boundary layer is a robust and generic response to sufficiently strong cross-equatorial pressure gradients. The transition does not require land-sea contrast, orography, or zonal asymmetries, but follows directly from large-scale meridional forcing. Alongside the increase in the advective terms, the latitude of the ITCZ also shifts northward in experiments with off-equatorial SST maxima (not shown). These results are consistent with previous studies highlighting the role of advective terms in determining the mean location of ITCZ(\citep{dixit2017role}).
 
\begin{figure}[!htb]
\centering
\noindent\includegraphics[width=\textwidth]{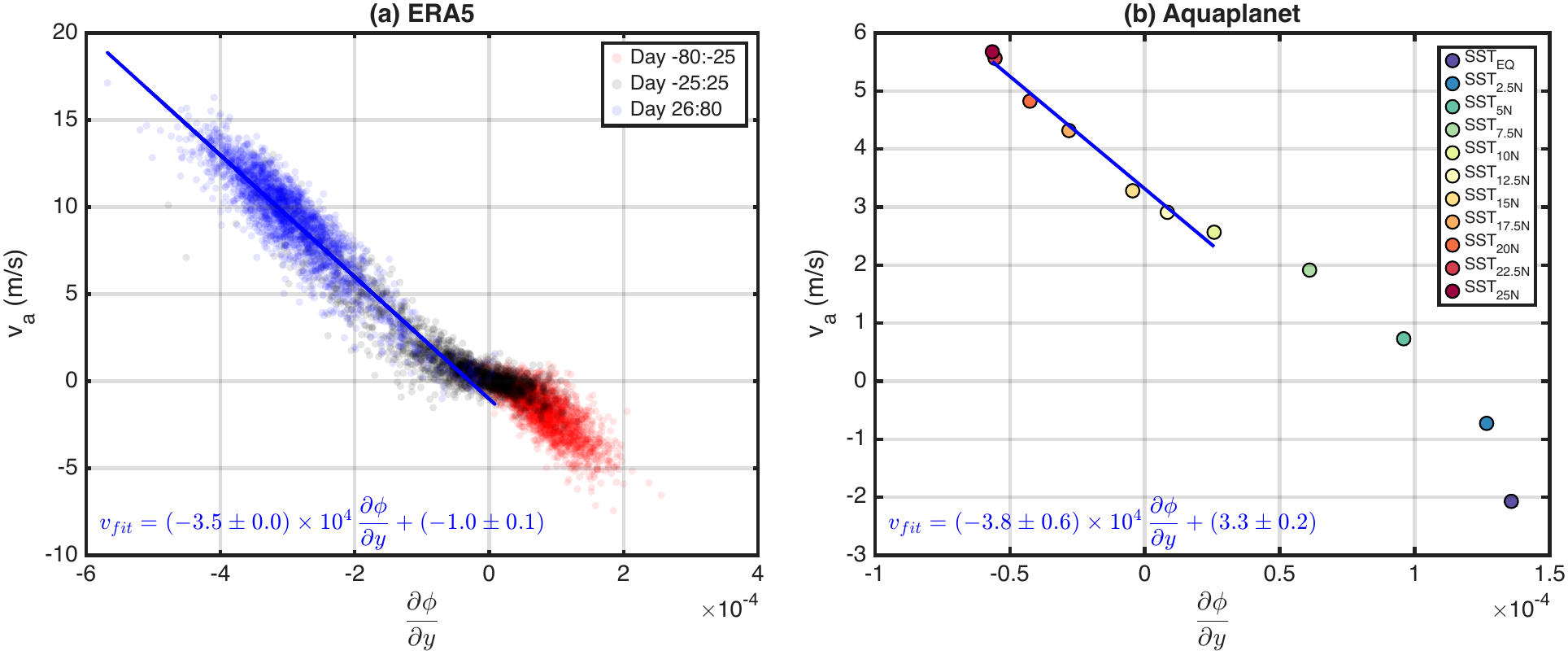}
\caption{(a) Scatter plot between daily averaged meridional geopotential gradient (in $m/s^2$) and  cross-isobaric meridional wind (in $m/s$) for composite periods based on ERA5 reanalysis data for 1979-2020. The least-square linear fit line after the development of ABL (day 26 to 80) is also plotted.
(b) Steady-state zonally averaged meridional geopotential gradient (in $m/s^2$) and cross-isobaric meridional wind (in $m/s$) averaged for the $0-10^\circ$N latitude zone, with different SST maximum experiments. The least-square linear fit line starting from experiment $\mathrm{SST_{10N}}$ till $\mathrm{SST_{25N}}$ is also plotted in blue. The equation of the fit is also indicated for both panels.}
\label{fig4}
\end{figure}

As discussed in Section \ref{3.2}, theory predicts that within the advective boundary layer the cross-isobaric meridional wind-speed scales linearly with the local meridional geopotential gradient, with a proportionality constant set by (i.e., equal to) the inertial timescale. This prediction is quantitatively supported by the aquaplanet simulations (Fig. \ref{fig4}b). Across experiments with strong cross-equatorial flow, the steady-state cross-isobaric meridional wind averaged over $0$–$10^\circ$N  is described well by a linear relationship of the form

\begin{equation}
	v_{a} = -(3.8\pm0.6)\times10^{4}\frac{\partial{\phi}}{\partial{y}} + (3.3\pm0.2)
 \label{eq5.17}
\end{equation}

where the fitted slope closely matches the inertial timescale at $\sim10^\circ$N and is consistent with observational estimates for the Somali jet. This linear scaling is most accurate in experiments where the SST maximum is displaced sufficiently north of the equator, resulting in a larger cross-equatorial pressure gradient, yielding a fully developed advective boundary layer (Fig. \ref{fig4}b). When the SST maximum lies closer to the equator, advective contributions are weaker and friction becomes significant in the zonal momentum budget (Fig. \ref{fig3}), leading to larger departures from the linear fit, as anticipated from the idealizations made in the theory above.

\subsubsection{Effects of changing the inertial timescale}

This section examines how the cross-equatorial flow and the advective boundary layer balances depend on the inertial timescale, which we vary by changing the planetary rotation rate in the aquaplanet simulations. This set of experiments tests the prediction that the sensitivity of cross-equatorial flow to geopotential gradients scales with the inverse Coriolis parameter.

\begin{figure}[!htb]
\centering
\noindent\includegraphics[width=\textwidth]{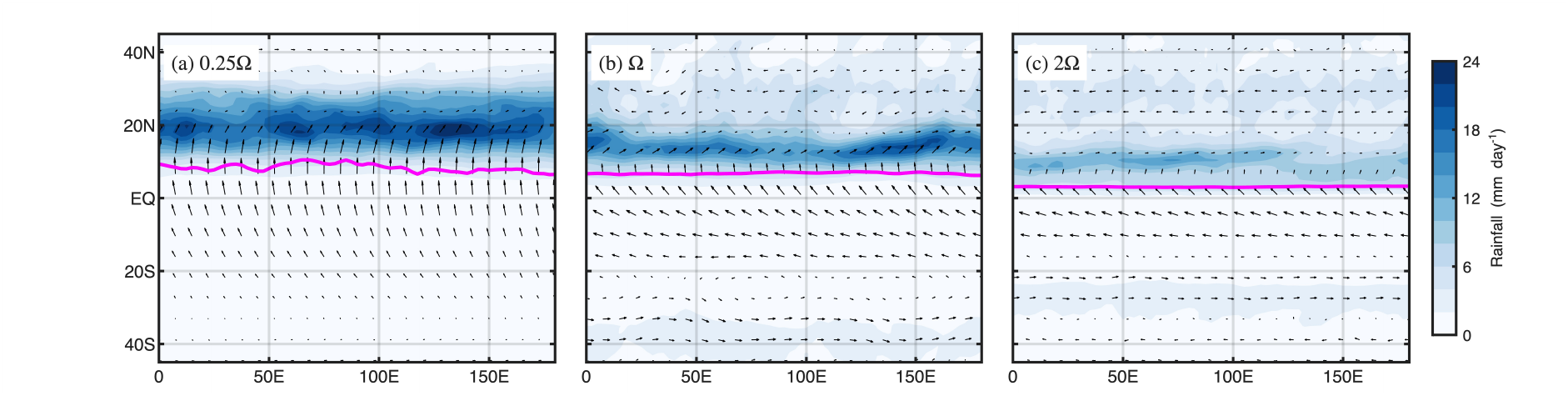}
\caption{Steady-state precipitation (shading in mm/day) for (a) $0.25\Omega$, (b) $\Omega$, and (c) $2\Omega$ planetary rotation rate for maximum SST located at 25$^\circ$N. Winds at 900 hPa and zero absolute vorticity ($\eta=0$) contour are also shown for each case.}
\label{fig5}
\end{figure}

We focus on simulations with the SST maximum fixed at $25^\circ$N ($\mathrm{SST_{25N}}$), and vary the rotation rate from $0.25\Omega$ to $2\Omega$ (current rotation rate of the Earth $\Omega=7.29 \times 10^{-5}$ rad/s). As planetary rotation rate is lowered, the inertial timescale increases, strengthening the cross-equatorial flow. In the $0.25\Omega$ case, the cross-equatorial flow is much stronger, and the zero absolute vorticity contour is displaced farther northward of the equator (Fig. \ref{fig5}a). Concomitantly, the region of high precipitation (ITCZ) broadens and shifts poleward, reflecting weaker rotational constraints on meridional motion. 

At the other extreme, for $2\Omega$, the cross-equatorial flow is weaker. After crossing the equator, the flow rapidly turns eastward under strong Coriolis deflections at a faster rotation rate, and the $\eta=0$ contour remains confined close to the equator. The ITCZ is narrower and more equatorward in this case (Fig. \ref{fig5}c).  This inverse relation between ITCZ latitude and planetary rotation rate is consistent with findings of \citet{faulk2017effects}.

\begin{figure}[!htb]
\centering
\noindent\includegraphics[width=\textwidth]{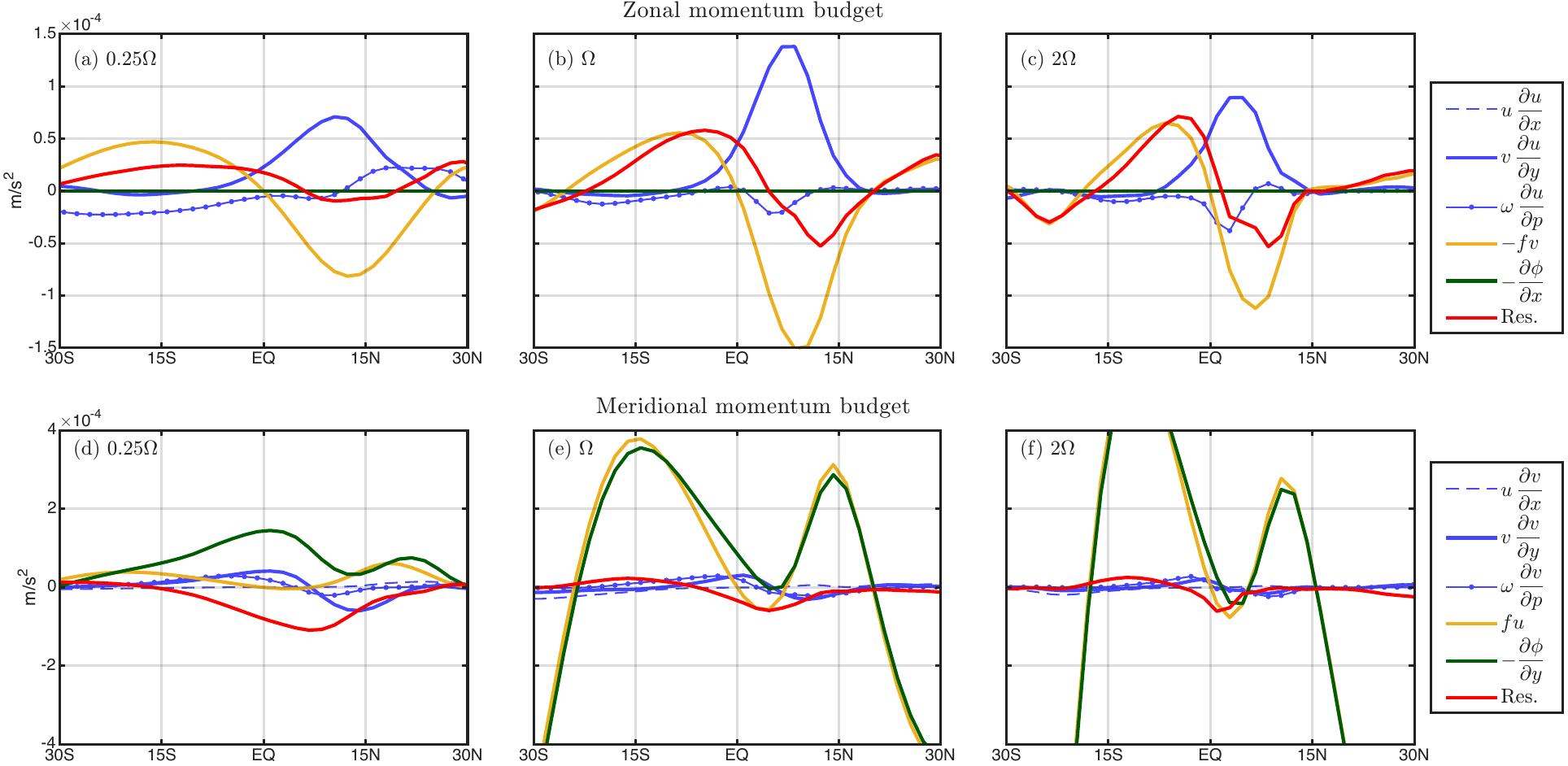}
\caption{Meridional profile of steady-state zonal (top row) and meridional momentum (bottom row) budget terms for $\mathrm{SST_{25N}}$ experiment for three different planetary rotation rates: (a,d) $0.25\Omega$, (b,e) $\Omega$, and (c,f) $2\Omega$.}
\label{fig6}
\end{figure}

Consistent with the stronger cross-equatorial flow, it is also seen that for slow rotation, the meridional advection of zonal momentum extends up to $20^\circ$N (Fig. \ref{fig6}a). This is in contrast with faster planetary rotation, where the meridional extent of the influence of this term in the tropics is much more confined near the equator (Fig. \ref{fig6}c). 
Interestingly, systematic changes in the meridional momentum balance are observed with these circulation differences. In the meridional momentum budget, faster rotation enforces geostrophic balance almost everywhere outside a narrow equatorial strip (Fig. \ref{fig6}f). In contrast, for the slow-rotation case, geostrophic balance breaks down over a wide latitude range extending from the equator to nearly $15^\circ$N, with the pressure gradient force increasingly balanced by friction (Fig. \ref{fig6}d). Planetary rotation rate also has consequences for the transition latitude (latitude where $u\approx v$) in these experiments, which increases for slow rotation. For the $0.25\Omega$ case, the transition latitude where meridional and zonal winds become equal is much farther northward (almost 10$^\circ$) than that for the $2\Omega$ experiment (Fig. S9).

\begin{figure}[!htb]
\centering
\noindent\includegraphics[width=\textwidth]{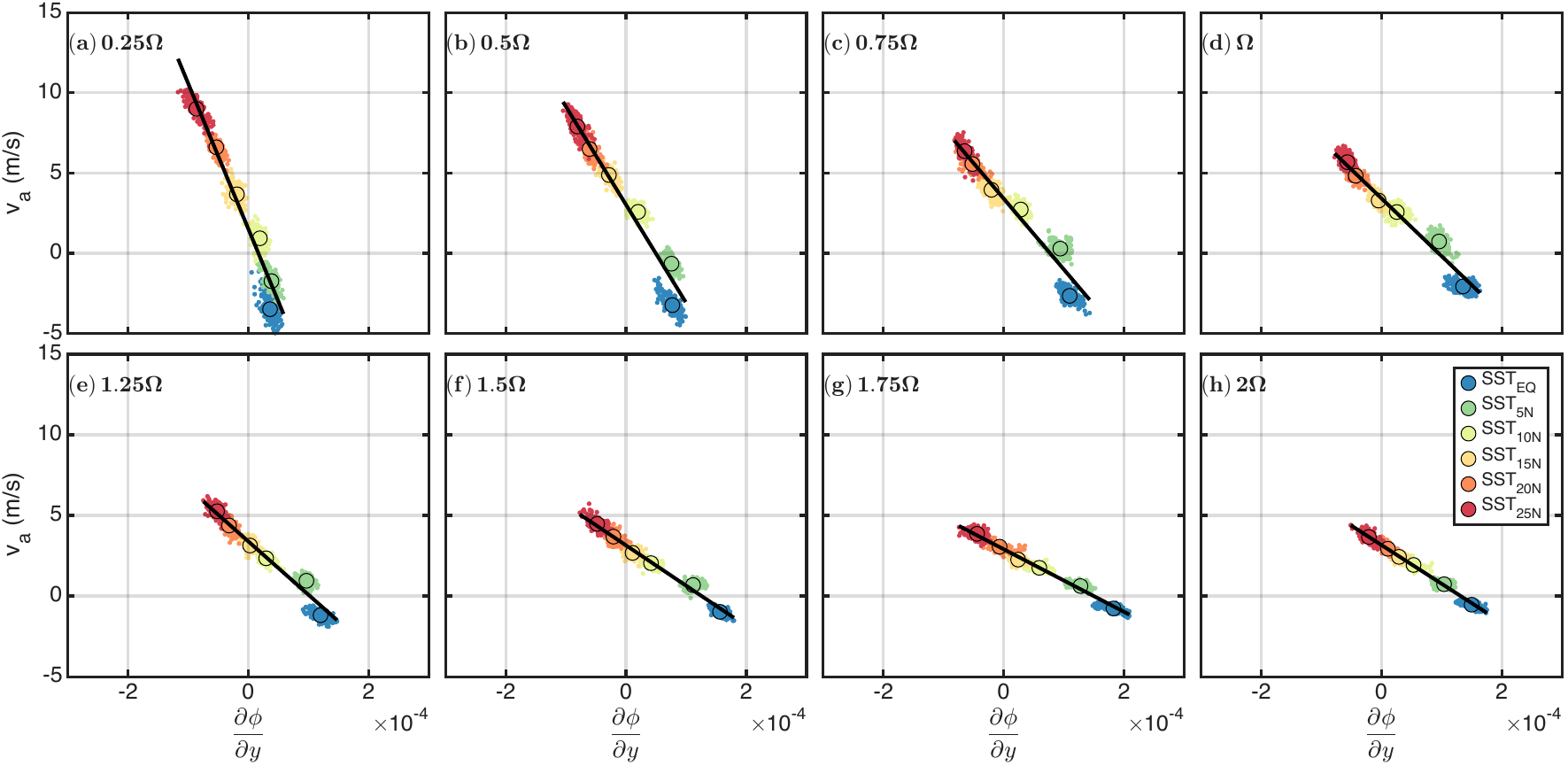}
\caption{Zonally averaged cross-equatorial meridional wind ($v_a$) and meridional geopotential gradient averaged from equator till the latitude nearest to the transition latitude ($y_{T}$ in table \ref{tb1}) for different SST maximum cases (represented with different marker color) for different planetary rotation rates (a-h). Small markers indicate the value on each individual day of the last 160 days of the simulation, and larger markers indicate the mean of all these days (steady-state values). The least-square linear fit for all the points is indicated by the black line in each panel. The coefficients of these linear fits are presented in table \ref{tb1}.
}
\label{fig7}
\end{figure}

We next examine how changes in the inertial timescale affect the relationship of proportionality and its coefficient between the meridional geopotential gradient ($\partial \phi/\partial y$) and the cross-isobaric meridional wind ($v_a$). As shown earlier \citep{masiwal2023explaining}, the product of these two terms plays a governing role in KE generation in the boundary layer. The aforementioned proportionality relation is also central to the observed quadratic dependence between KE generation and meridional geopotential gradients during the seasonal transition of the Somali jet and the South Asian monsoon \citep{masiwal2023explaining}. The above results, such as Fig. \ref{fig4} and equation \ref{eq5.17}, already suggest that for an aquaplanet with Earth's rotation, the dependence of $v_{a}$ on $\frac{\partial{\phi}}{\partial{y}}$ is similar to that observed in Earth observations. It is also evident that the approximately linear dependence of $v_a$ on $\partial \phi/\partial y$ identified for Earth-like rotation persists across all rotation rates examined (Fig. \ref{fig7}). Specifically, slower rotation leads to a larger slope of the linear relationship between $v_a$ and $\partial \phi/\partial y$, indicating that a given pressure gradient produces a larger cross-isobaric flow. Faster rotation produces the opposite effect. This behavior is qualitatively consistent with the theoretical prediction that the proportionality constant scales with the inertial timescale ($\sim 1/f$). 

\begin{figure}[!htb]
\centering
\noindent\includegraphics[width=\textwidth]{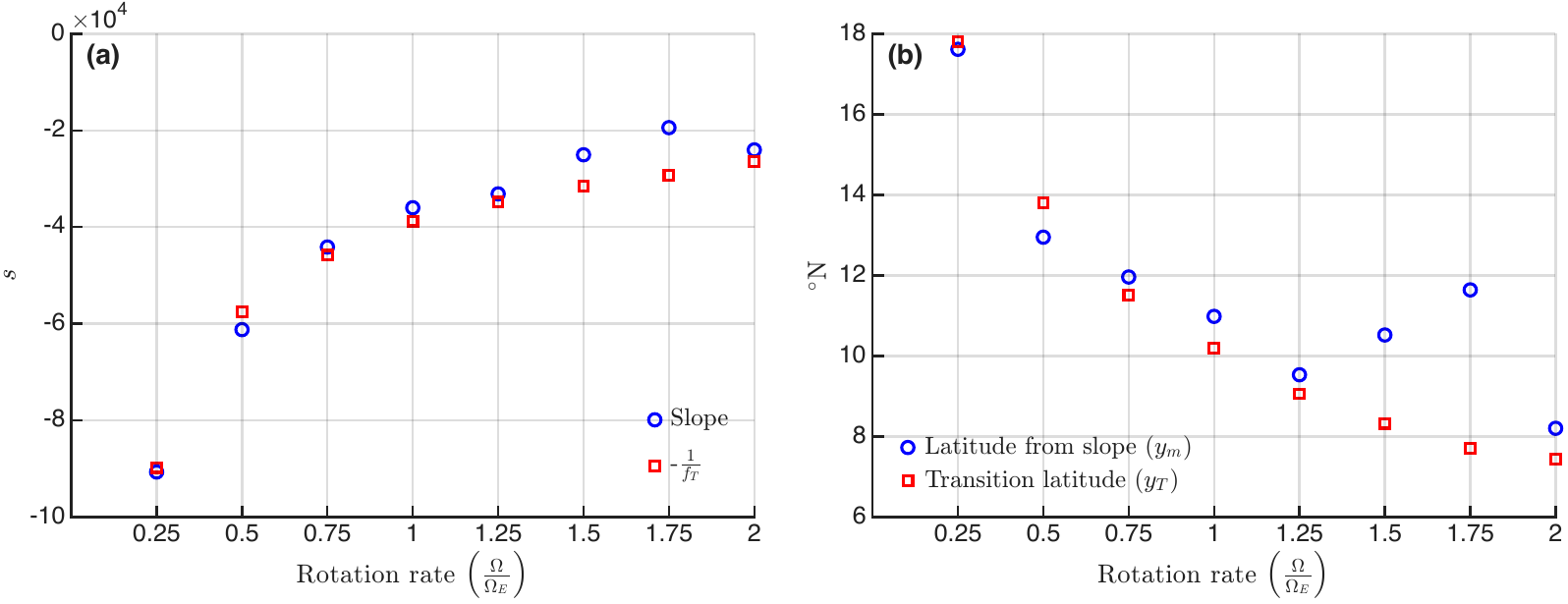}
\caption{(a) Variation of the slope (blue circles), estimated from a least-squares linear fit in Fig. \ref{fig7}, for different planetary rotation rates. The red squares indicate the theoretical scaling $-1/f_T$, where $f_T$ is the Coriolis parameter evaluated at the transition latitude $y_T$ reported in Table \ref{tb1}. (b) Comparison between the latitude inferred from the fitted slope, $y_m$ (blue circles), and the diagnosed transition latitude $y_T$ (red squares) for different planetary rotation rates.}
\label{fig8}
\end{figure}

To quantify this dependence, we perform least-squares linear fits using daily-mean data from the final 160 days of each simulation, pooling all SST experiments for a given rotation rate. The resulting slopes and intercepts are summarized in Table \ref{tb1}. While the fitted slopes decrease monotonically with increasing rotation rate, their magnitudes do not match exactly with the simple theoretical prediction $\sim 1/f$ quantitatively, especially for $1.5\Omega$ and $1.75\Omega$ experiments (Fig. \ref{fig8}a). Consequently, the latitude inferred from the fitted slope for these two cases does not decrease monotonically, contrary to the expected decrease in transition latitude with increasing rotation rate, as the inertial timescale shortens (Fig. \ref{fig8}b). These discrepancies likely reflect violations of idealized assumptions, including neglecting the frictional term and the coarse resolution of our model simulations, which is likely insufficient to resolve the key dynamics of latitude close to the equator at higher rotation. Past studies have mentioned the key role of resolution in accurately quantifying the contribution of advective terms in boundary layer momentum balance~\citep{gonzalez2019violation,masiwal2023explaining,praturi2025meridional}. Additionally, the intercept of the linear fit has no direct analogue in the idealized theoretical model. Nonetheless, despite its simplicity, the scaling captures the direction and qualitative dependence of the sensitivity of cross-equatorial flow on the inertial timescale (Fig. \ref{fig8}a), lending support to the proposed advective boundary layer framework.

\begin{table}[!htb]
\centering

\begin{tabular}{@{}ccccc@{}}
\toprule
Rotation rate & Slope $(\times 10^4)$ & Intercept & Transition Latitude ($y_T$ in $^\circ $N) & $-1/f_T$ $(\times 10^4)$ \\ 
\midrule
$0.25\Omega$ & $-9.1 \pm 0.1$ & $1.6 \pm 0.13$ & 17.8 & $-9.0$ \\
$0.5\Omega$  & $-6.1 \pm 0.09$ & $3.0 \pm 0.06$ & 13.8 & $-5.7$ \\
$0.75\Omega$ & $-4.4 \pm 0.08$ & $3.6 \pm 0.05$ & 11.5 & $-4.5$ \\
$\Omega$     & $-3.6 \pm 0.05$ & $3.4 \pm 0.04$ & 10.2 & $-3.9$ \\
$1.25\Omega$ & $-3.3 \pm 0.05$ & $3.0 \pm 0.04$ & 9.1  & $-3.5$ \\
$1.5\Omega$  & $-2.5 \pm 0.03$ & $2.8 \pm 0.03$ & 8.3  & $-3.1$ \\
$1.75\Omega$ & $-1.9 \pm 0.02$ & $2.5 \pm 0.02$ & 7.7  & $-2.9$ \\
$2\Omega$    & $-2.4 \pm 0.02$ & $2.3 \pm 0.02$ & 7.4  & $-2.6$ \\
\bottomrule

\end{tabular}
\caption{Slope and intercept estimated from a least-squares linear fit of $v_a$ for different planetary rotation rates. The linear fit and the data used to estimate this fit are shown in Fig.~\ref{fig7}. The transition latitude ($y_{T}$, where $u\approx v$) is obtained from the steady-state meridional profile of $u$ and $v$ from $\mathrm{SST_{25N}}$ experiment for each case. The inertial timescale corresponding to this transition latitude ($1/f_T$) is also listed. }
\label{tb1}

\end{table}

\section{Summary and Discussion} \label{5.4}
In this study, we examined the structure, dynamics, and emergence of the advective boundary layer (ABL) in the tropical atmosphere, focusing on the cross-equatorial flow that develops into the Somali jet, during South Asian monsoon onset. Analysis of the boundary-layer momentum budget in the ``N.Eq.'' region (50$^\circ$E-60$^\circ$E, 2$^\circ$N-10$^\circ$N) reveals a clear regime transition at the onset of persistent cross-equatorial flow: the zonal momentum balance shifts from a frictionally dominated Ekman regime to one governed by meridional advection of zonal momentum (Fig. \ref{fig1}a,c). This transition coincides with increasing local Rossby number ($R_o$) and a reduction of absolute vorticity towards zero (Fig. S2), signaling breakdown of Ekman balance. In contrast, the meridional momentum balance remains close to geostrophic throughout the transition (Fig.~\ref{fig1}b,d), highlighting a fundamental asymmetry between zonal and meridional momentum balances. This asymmetry arises from the strong anisotropy of horizontal length scales, with sharp meridional gradients dominating the relative vorticity. The local Rossby number therefore offers a natural diagnostic of this regime shift, as it captures the relative importance of advective accelerations relative to Coriolis accelerations and reflects the scale separation between meridional and zonal variations in the boundary layer flow. This scale difference also underlies the observed contrast between a geostrophic balance in the meridional momentum equation, and a transition to an advective (rather than frictional) balance in the zonal momentum. The local Rossby number thus characterizes both the evolving dynamical regime and the background anisotropy that gives rise to the ABL.

To explain this transition mechanistically, we developed a scaling theory based on dimensional analysis of the horizontal momentum equations. As boreal summer insolation increases and northern hemispheric sea surface temperatures rise, the meridional length scales of both the geopotential field and zonal winds contract. When the product of these scales approaches $\phi/f^2$, absolute vorticity in the boundary layer vanishes and the flow enters the advective boundary layer regime (figure \ref{fig2}). This regime persists through the monsoon season, while expansion of these length scales during withdrawal restores the frictionally dominated Ekman flow. 

In the ABL regime, where $\eta\rightarrow0$, the boundary-layer kinetic energy balance simplifies: generation by pressure work is balanced primarily by advection. This balance motivates a simple linear kinetic-energy model with an approximately constant advective decay timescale (equation \ref{eq5.13}). The resulting diagnostic relation between meridional geopotential gradient and meridional wind speed emerges directly, with a proportionality set by the advective time scale. Observations from the Somali jet region show that this timescale is well approximated by the inertial timescale ($1/f$) near the transition latitude, lending support to our proposed theory.

The robustness of this theory was tested using a suite of moist aquaplanet GCM experiments. By varying the latitude of the imposed SST maximum, we systematically altered cross-equatorial pressure gradients and ITCZ position in the zonally symmetric experiments. As SST maxima were displaced northward, cross-equatorial flow intensified, the zero-absolute-vorticity contour migrated poleward, and the advective boundary layer expanded accordingly, concomitantly with increase in meridional advection of zonal momentum. Across these experiments, the meridional momentum balance remained close to geostrophic, while the zonal momentum balance transitioned to advective control (Fig. \ref{fig3}).

Crucially, the linear diagnostic relation between cross-isobaric meridional wind and meridional geopotential gradient remained valid across a wide range of SST configurations, with a constant of proportionality close to that inferred from observations as well as derived from theory (Fig.~\ref{fig4}). Varying the planetary rotation rate further demonstrated that slower rotation enhances cross equatorial flow and shifts the transition latitude poleward (Fig.~\ref{fig5}), as well as amplifies the sensitivity of meridional winds to geopotential gradient in accordance with the increasing inertial timescale (Fig.~\ref{fig7}, Table~\ref{tb1}).  At very slow rotation, deviations from geostrophic balance in the meridional momentum equation introduce secondary frictional effects, modifying but not eliminating—the advective regime and the sensitivity of meridional winds to geopotential gradients.

These results are consistent with earlier work documenting the breakdown of Ekman balance and the growing importance of advective dynamics in off-equatorial convergence zones \citep{tomas1999influence,schneider2008eddy, gonzalez2016dynamics,faulk2017effects}. Some of these studies—particularly those using dry GCMs—suggest that the emergence of advective dynamics does not strictly require moist processes. Nevertheless, the role of moist convection on the transition, maintenance and potential destabilization of the ABL remains an important open question. Observational analyses in the East Pacific similarly link strong cross-equatorial pressure gradients and negative absolute vorticity to ITCZ variability \citep{toma2010oscillations, gonzalez2022rapid}. Absolute vorticity in the boundary layer is also known to influence the tropical intraseasonal variability during the boreal summer~\citep{masiwal2025intraseasonal}. Given the close association between absolute vorticity and ABL, it is important to examine how the emergence of ABL modulates monsoon rainfall, intraseasonal variability, and predictability over South Asia. Likewise, the physical drivers of meridional length-scale contraction in geopotential and wind fields during seasonal transitions warrant further investigation.

In summary, this study provides a mechanistic framework for understanding the tropical boundary layer’s transition to an advective regime, grounded in new scaling theory and supported by reanalysis and idealized modeling. The results clarify how pressure gradients, horizontal length scales, and planetary rotation jointly determine boundary-layer momentum balances, explaining the recurrent emergence of the ABL during South Asian monsoon onset. More broadly, this framework offers a benchmark for improving representation of low-latitude boundary layer dynamics in weather and climate models, particularly in regimes where frictional closures are inadequate. Future work will explore the external controls on ABL emergence and assess its role in shaping tropical variability across seasonal and intraseasonal timescales.

\section*{Acknowledgments}
The authors thank Gilles Bellon for helpful discussion. RM acknowledges the assistance of Jerry B. Samuel and Abu Bakar Siddiqui Thakur in porting the model to the local computing cluster. 

\datastatement
Reanalysis data used in this study from ERA5 is publicly available from Copernicus Climate Change Service and can be accessed at \url{https://cds.climate.copernicus.eu/datasets/reanalysis-era5-pressure-levels?tab=overview}. The India Meteorological Department (IMD) 1-degree gridded rainfall data is available at \url{https://www.imdpune.gov.in/cmpg/Griddata/Rainfall_1_NetCDF.html}. The outputs of the model experiments and the analysis code can be made available upon request.

\bibliographystyle{ametsocV6}
\bibliography{sample}

\renewcommand{\thefigure}{S\arabic{figure}} 





\Large{\textbf{Supplementary for} "Emergence of an Advective Boundary Layer in Monsoon Cross-Equatorial Flow: Scaling, Dynamics, and Idealized Models"}

\authors{Rajat Masiwal,\aff{a,b}\correspondingauthor{Rajat Masiwal, rajatmasiwal@iisc.ac.in} 
Ashwin K Seshadri,\aff{b,c}
	Vishal Dixit,\aff{d}  
}

\affiliation{
    \aff{a}{Department of the Geophysical Sciences, The University of Chicago, Chicago, IL, USA}\\
    \aff{b}{Centre for Atmospheric and Oceanic Sciences, Indian Institute of Science, Bengaluru, India}\\
	\aff{c}{Divecha Centre for Climate Change, Indian Institute of Science, Bengaluru, India}\\
	\aff{d}{Centre for Climate Studies, Indian Institute of Technology Bombay, Mumbai, India}\\
}
%


\maketitle

\begin{itemize}
    \item Supplementary Figures S1-S9
\end{itemize}

\clearpage

\setcounter{figure}{0}

\begin{figure}[!htb]
\centering
\noindent\includegraphics[width=\textwidth]{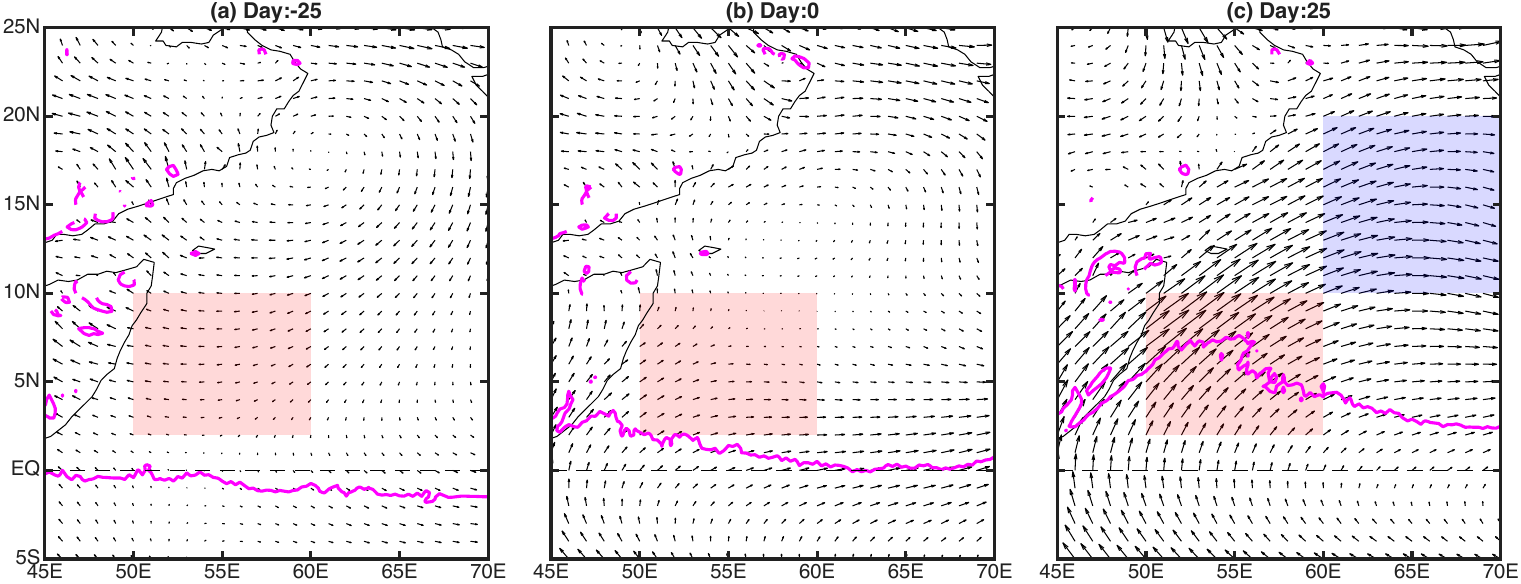}
\caption{Composites of 900 hPa horizontal winds and zero absolute vorticity ($\eta$) contour (magenta line) for (a) 25 day before, (b) on the day, and (c) 25 day after the onset of cross-equatorial flow. The red box indicates the ``N.Eq.'' region (50$^\circ$E-60$^\circ$E, 2$^\circ$N-10$^\circ$N) used for averaging in the main text. The blue box in (c) indicates the region where the jet is zonal (60$^\circ$E-70$^\circ$E, 10$^\circ$N-20$^\circ$N).}
\label{sfig1}
\end{figure}

\begin{figure}[!htb]
\centering
\noindent\includegraphics[width=\textwidth]{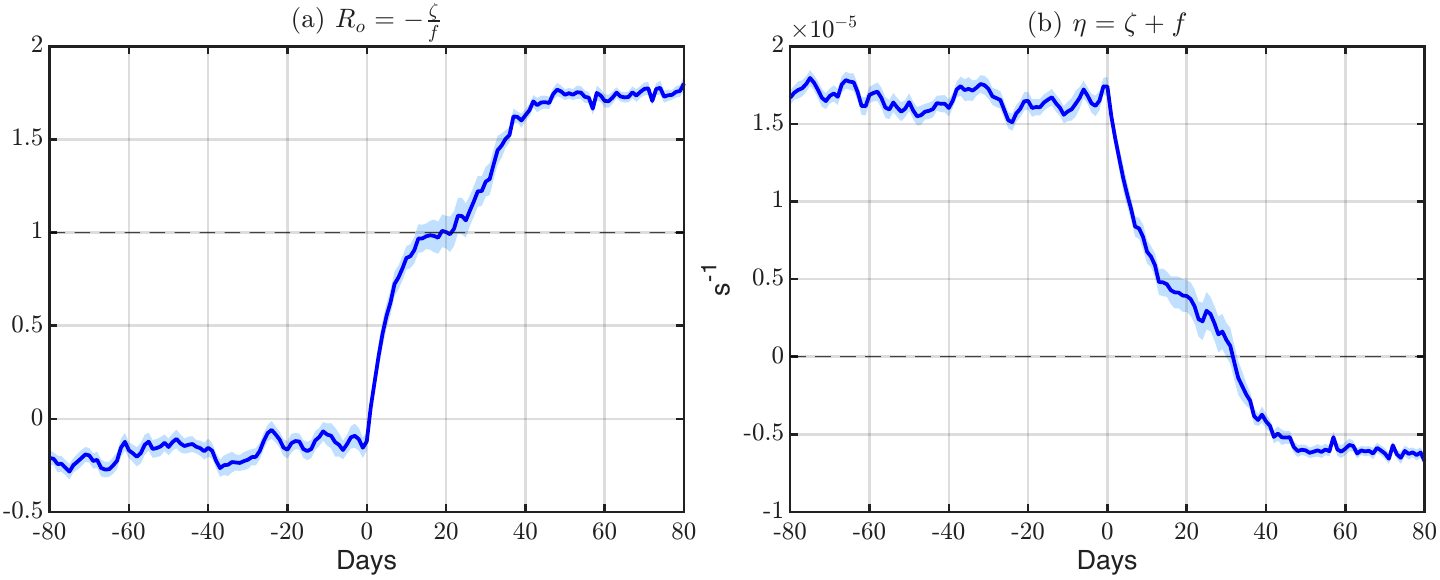}
\caption{Composite evolution of 900 hPa (a) local Rossby number ($R_{o}$) and (b) absolute vorticity ($\eta$ in $s^{-1}$). Both the terms are averaged over the ``N.Eq.'' region}
\label{sfig2}
\end{figure}

\begin{figure}[!htb]
\centering
\noindent\includegraphics[width=0.65\textwidth]{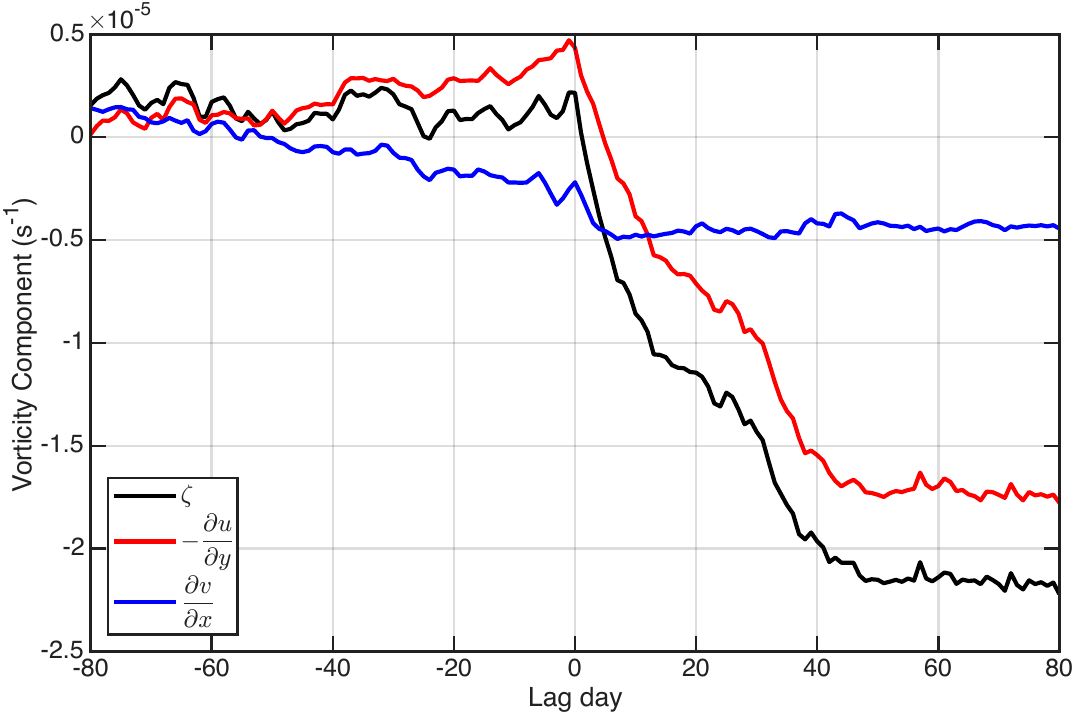}
\caption{Composite evolution of 900 hPa relative vorticity ($\zeta$) and its two components. }
\label{sfig9}
\end{figure}

\begin{figure}[!htb]
\centering
\noindent\includegraphics[scale=0.65]{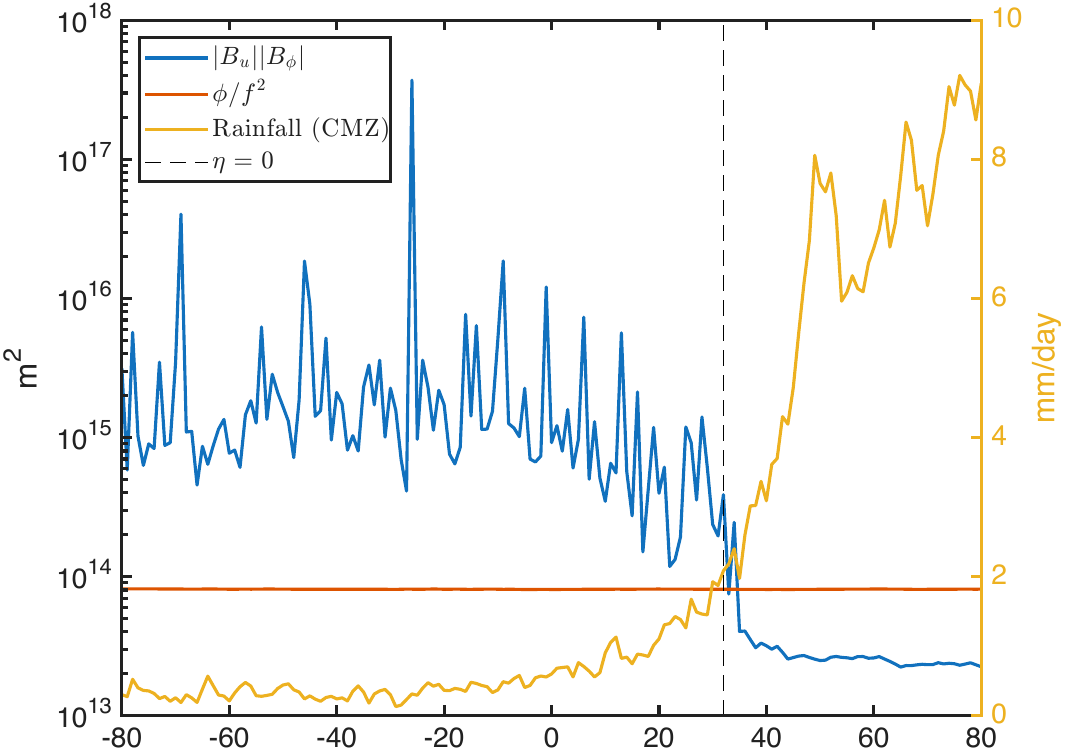}
\caption{Composite evolution of $B_{\phi}B_{u}$ and $\phi/f^2$ for the period of 1979 to 2020 averaged over the ``N.Eq.'' region. The length scales are estimated using an exponential fit. The vertical dashed line indicates the day when $\eta$ changes sign from positive to negative. The yellow line represents the composite evolution of daily rainfall (in $mm/day$) averaged over the core monsoon zone (CMZ) based on India Meteorological Department (IMD) data.}
\label{sfig3}
\end{figure}

\begin{figure}[!htb]
\centering
\noindent\includegraphics[width=\textwidth]{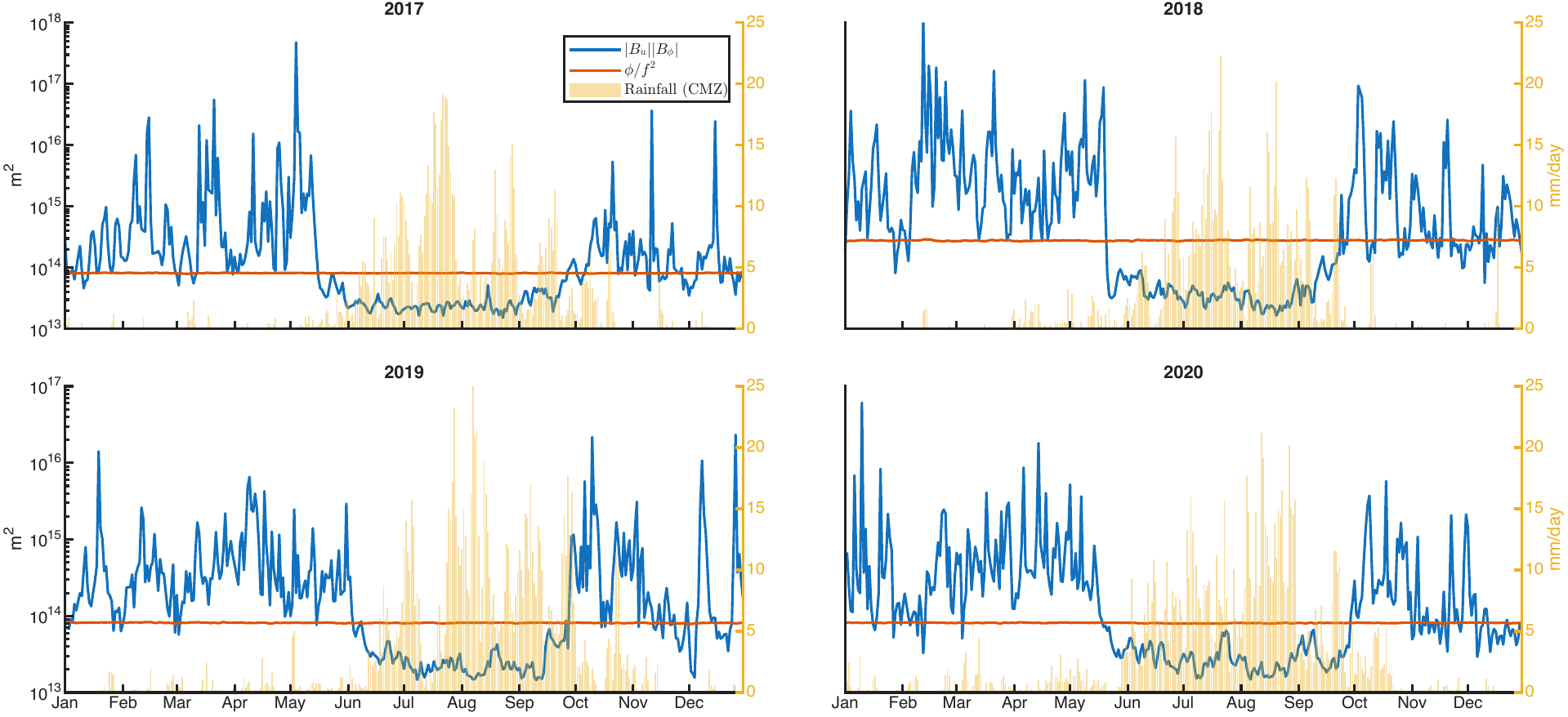}
\caption{Evolution of daily of $B_{\phi}B_{u}$ and $\phi/f^2$ over the ``N.Eq.'' region for 2017 to 2020. The length scales are estimated using an exponential fit. The bars represent the daily rainfall (in $mm/day$) averaged over the core monsoon zone (CMZ) based on India Meteorological Department (IMD) data.}
\label{sfig4}
\end{figure}

\begin{figure}[!htb]
\centering
\noindent\includegraphics[width = \textwidth]{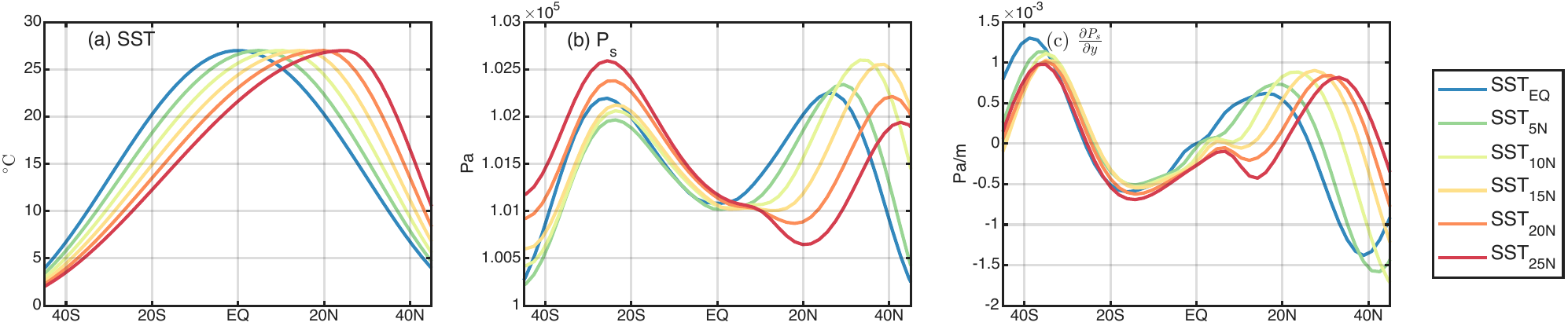}
\caption{(a) Meridional profile of prescribed SST for different experiments, based on equation 2 in the main text, with varying $y_{0}$. These SST profiles impose distinct meridional profiles of steady-state (b) surface pressure (P$_S$) and (c) meridional pressure gradient.}
\label{sfig5}
\end{figure}

\begin{figure}[!htb]
\centering
\noindent\includegraphics[width = \textwidth]{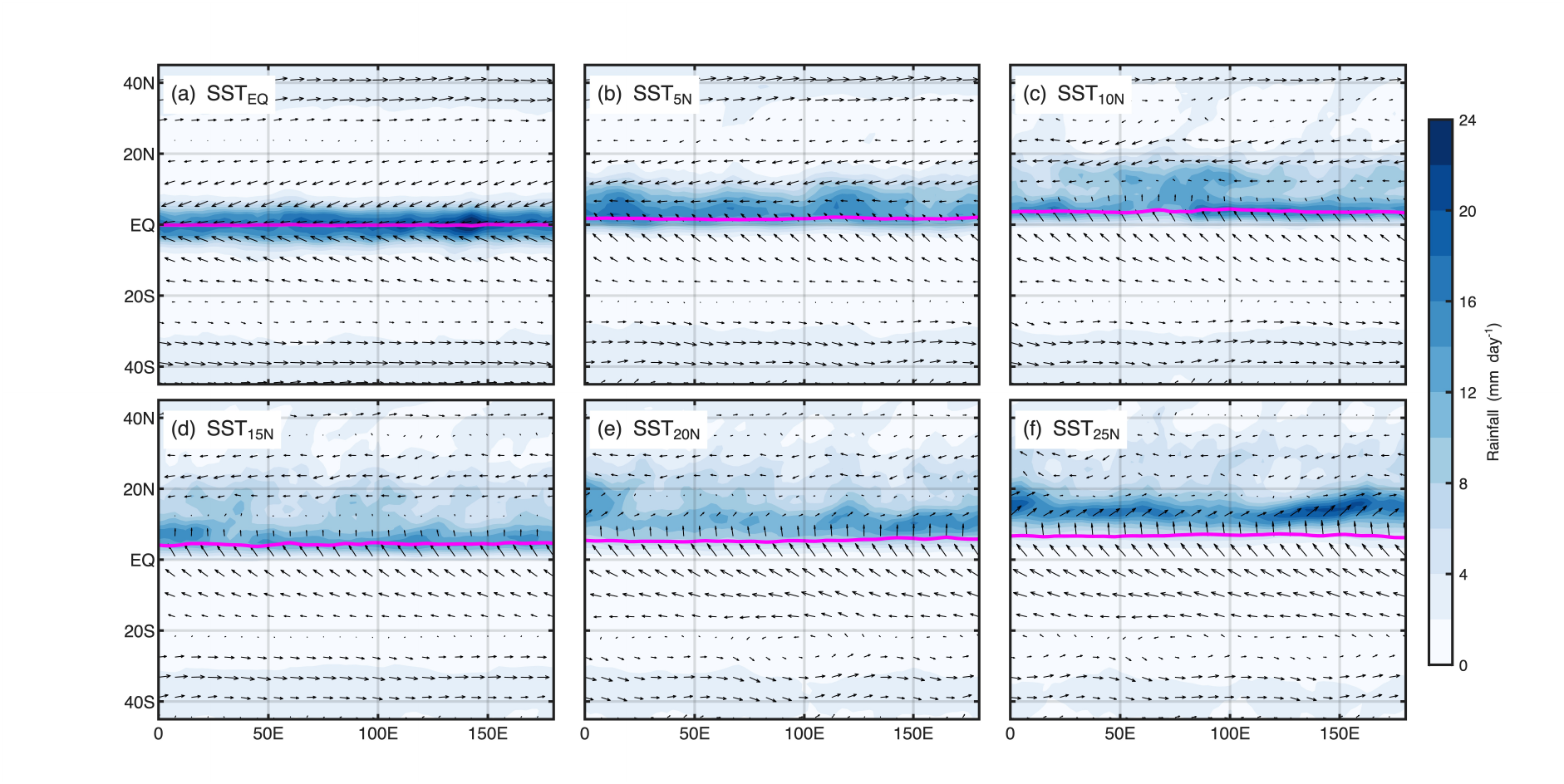}
\caption{Steady-state precipitation (shading in mm/day) for different SST maximum experiments. Winds at 900 hPa and zero absolute vorticity ($\eta=0$) contour are also shown for each case.}
\label{sfig6}
\end{figure}

\begin{figure}[!htb]
\centering
\noindent\includegraphics[width = \textwidth]{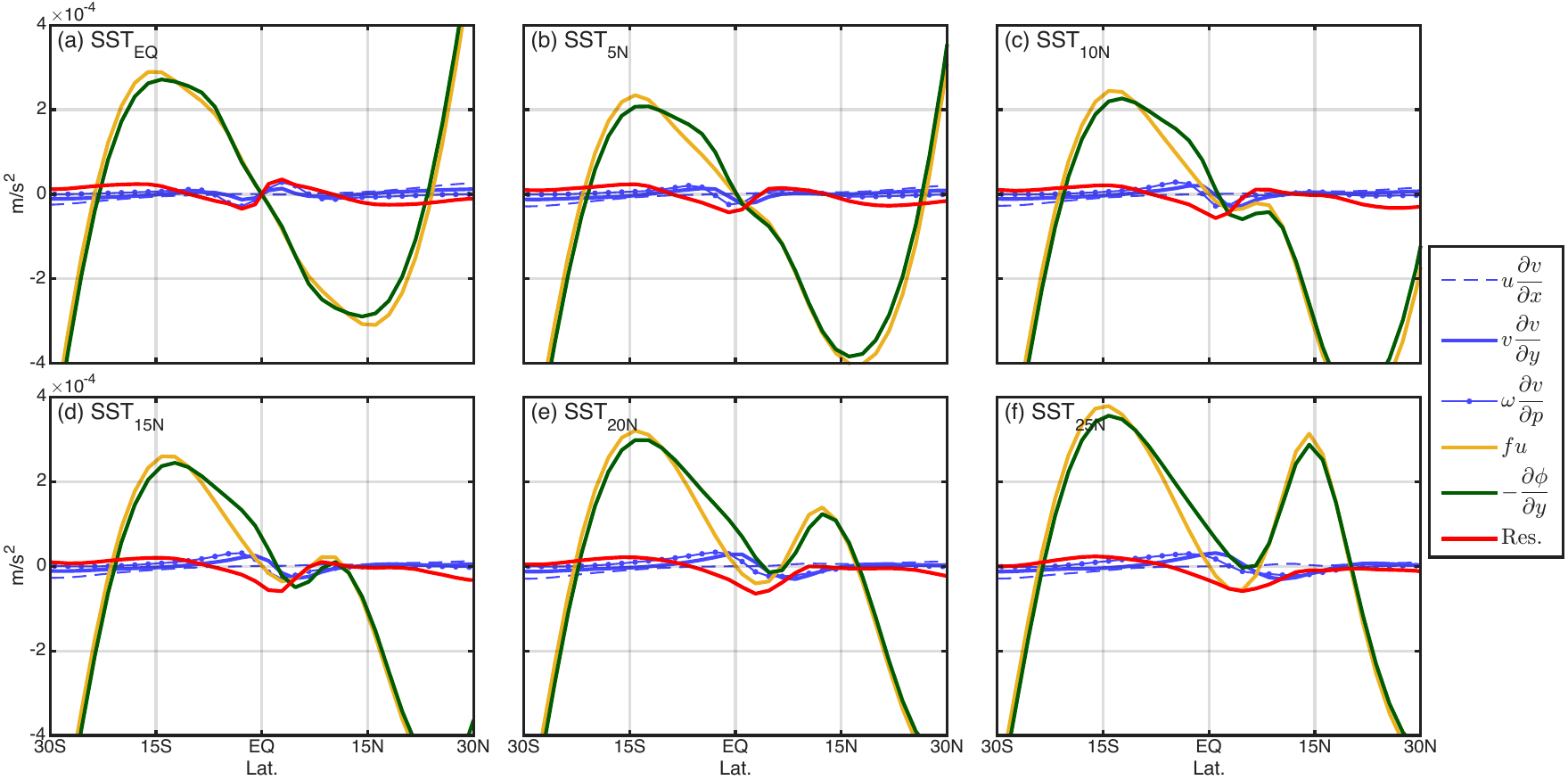}
\caption{Meridional variation of meridional momentum terms for different SST maximum experiments.}
\label{sfig7}
\end{figure}

\begin{figure}[!htb]
\centering
\noindent\includegraphics[width = \textwidth]{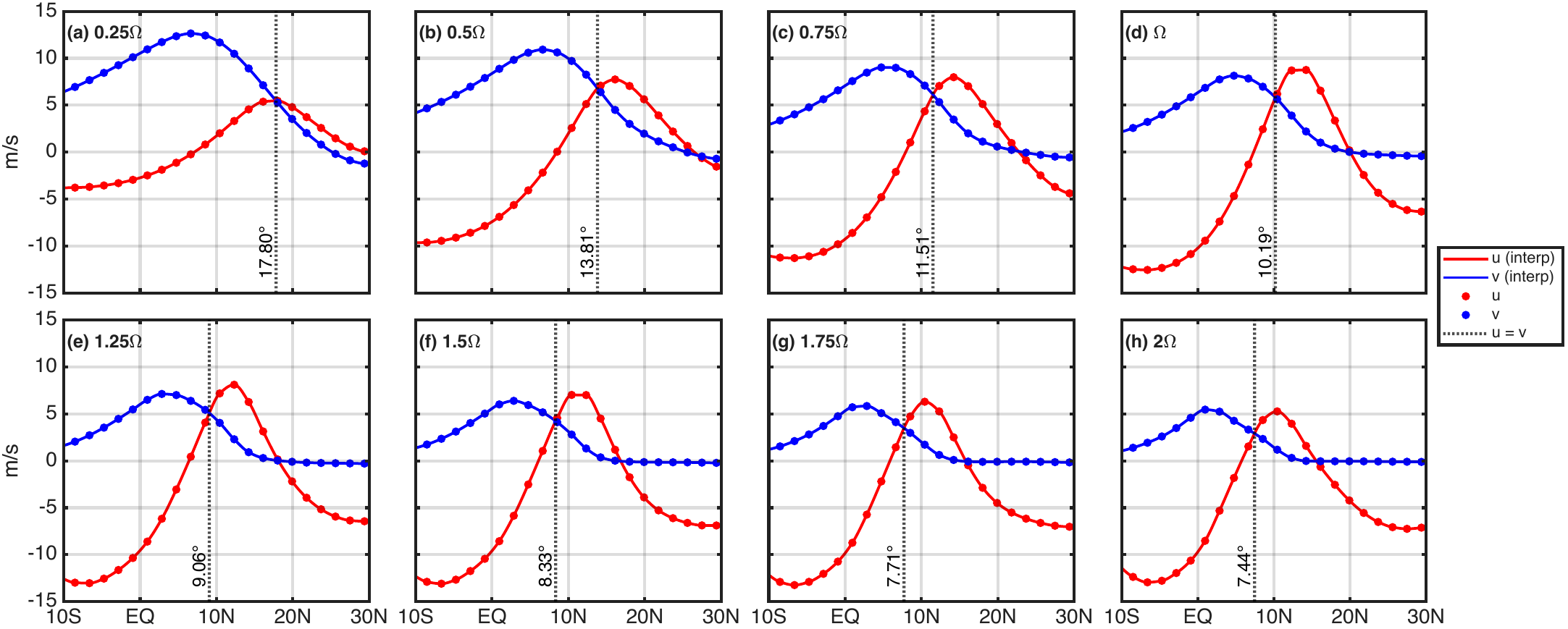}
\caption{Latitude profiles of steady-state zonal and meridional winds in different rotation experiments for $\mathrm{SST_{25N}}$ case. Red and blue lines show the interpolated (using piecewise cubic Hermite interpolation) zonal (u) and meridional (v) wind profiles, respectively, while circles indicate the values at model's native resolution. Vertical dotted lines mark the transition latitude ($u\approx v$) computed from the interpolated wind profiles for each experiment, with the corresponding latitude value labeled near the x-axis.}
\label{sfig8}
\end{figure}


\end{document}